\documentstyle[12pt,dina4p]{article}
\newcommand{\be}{\begin{equation}}
\newcommand{\ee}{\end{equation}}
\newcommand{\ba}{\begin{eqnarray}}
\newcommand{\ea}{\end{eqnarray}}
\newcommand{\bc}{\begin{center}}
\newcommand{\ec}{\end{center}}

\newcommand{\bay}{\begin{array}{rcl}}
\newcommand{\eay}{\end{array}}
\newcommand{\dis}{\displaystyle}
\newcommand{\text}{\textstyle}

\newcommand{\rf}[1]{(\ref{#1})}
\newcommand{\R}{{I\!\! R}}
\newcommand{\Sp}{ \mbox{{\it Sp}$(2N)$}}
\newcommand{\spl}{\mbox{{\it sp}$(2N)$}}
\newcommand{\Mp}{ \mbox{{\it Mp}$(2N)$}}
\newcommand{\mpl}{\mbox{{\it mp}$(2N)$}}
\newcommand{\rcl}{{\rm cl}}
\newcommand{\ra}{\rangle}
\newcommand{\la}{\langle}
\newcommand{\tfrac}[2]{{\text\frac{#1}{#2}}}
\newcommand{\da}{\dagger}

\newcommand{\omcirc}{\stackrel{\circ}{\omega}{}\kern -.3em}

\newcommand{\bel}[1]{\begin{equation}\label{#1}}
\renewcommand{\arraystretch}{2.0} 

\renewcommand{\hat}{\widehat} 
\renewcommand{\tilde}{\widetilde} 
  \renewcommand{\theequation}{\thesection.\arabic{equation}}
  \newcommand{\mysection}[1]{\section{#1}\setcounter{figure}{0}
                    \setcounter{table}{0}\setcounter{equation}{0}}

\def\DD{D\kern-0.82em\lower-.2ex\hbox{$\not$}\kern+0.8em}

\newcommand{\ccup}[1]{
   \mbox{$ \bigcup \kern -0.75em  \lower .9ex \hbox{${}_{#1}$}$}
     \kern +.5em
       }

\def\fs{
   \mbox{$ F_{\mu\nu}^a \kern -1.65em  \lower -.2ex \hbox{${}^*$}$}
     \kern +1.5em
       }
\def\fsn{
   \mbox{$ F \kern -0.9em  \lower -.2ex \hbox{${}^*$}$}
     \kern +0.35em
       }

\def\cc{\mbox{${\cal C}$}}
\def\cd{\mbox{${\cal D}$}}

\def\cg{\mbox{${\cal G}$}}
\def\ch{\mbox{${\cal H}$}}

\def\ck{\mbox{${\cal K}$}}
\def\cl{\mbox{${\cal L}$}}
\def\cm{\mbox{${\cal M}$}}
\def\cn{\mbox{${\cal N}$}}
\def\co{\mbox{${\cal O}$}}
\def\cp{\mbox{${\cal P}$}}

\def\cv{\mbox{${\cal V}$}}

\begin{document}
\noindent
\begin{titlepage}
\begin{flushright}
DESY 97--127\\
hep-th/9804036
\end{flushright}
  \vspace{15mm} {\LARGE
  \begin{center}
    {\bf Quantum Mechanics as a  
      Gauge Theory 
       \\of Metaplectic Spinor Fields
\footnote{To appear in Int. J. Mod. Phys. A.}
  \vspace{0.2in}
     } 
\\ \vspace{0.3in}
  \end{center}}
  \vspace{0.2in} 
  \begin{center}
    {\large \sc {M. Reuter}} \\[.3cm] 
     \normalsize
     \it {Deutsches Elektronen-Synchrotron DESY\\
     Notkestrasse 85,
     D-22603 Hamburg,
     Germany
\footnote{Present address: Institut f\"ur Physik, Universit\"at Mainz,
                           D-55099 Mainz, Germany}
     }\\ 
  \end{center}
\vspace{0.3in}
 \normalsize

\begin{abstract}
A hidden gauge theory structure of quantum mechanics which
is invisible in its conventional formulation is uncovered.
Quantum mechanics is shown to be equivalent to a certain
Yang--Mills theory with an infinite--dimensional gauge group
and a nondynamical connection. It is defined over an arbitrary
symplectic manifold which constitutes the phase--space of the
system under consideration. The ''matter fields'' are local
generalizations of states and observables; they assume
values in a family of local Hilbert spaces (and their tensor products)
which are attached to the points of phase--space. 
Under local frame rotations they transform
in the spinor representation
of the metaplectic group \Mp, the double covering of \Sp. The
rules of canonical quantization are replaced by two independent
postulates with a simple group theoretical and differential
geometrical interpretation. A novel background--quantum split
symmetry plays a central r\^ole.
\end{abstract}
\end{titlepage}

\mysection{Introduction}
Both in general relativity and in Yang--Mills theory the principle
of local gauge invariance plays a central r\^ole. At the classical
level, these theories are based upon local structures such as the 
tangent spaces to the space--time manifold or local \linebreak
''color'' spaces,
as well as on connections (gauge fields) which provide a link
between such spaces sitting at different points of space--time. However,
at the quantum level, the importance of such local concepts is
considerably 
reduced. One has to introduce nonlocal objects which
are not related to any specific point of space--time, and which have
no natural interpretation within the classical theory of fiber
bundles. The most important object of this kind   is the Hilbert
space of physical states. According to standard canonical quantization,
every quantum system 
is described
by a single ''global'' Hilbert space whose elements (wave functions)
give rise to a probability density over configuration--space. It is one
and the same Hilbert space which governs the probabilities at all
points of configuration space and it is not possible
to associate this Hilbert space to a specific point of configuration
space. 

Obviously there
is a remarkable conceptual clash between the classical geometry of the 
field theories employed in particle physics and gravity on one hand 
and the standard formulations of quantum mechanics on the other.
In the former case one is dealing with fiber bundles over space--time.
The matter fields are ''living'' in certain vector spaces (''fibers'')
which are erected over each point of space--time. Vector fields, say,
assume values in the tangent spaces of the base manifold, or matter
fields carrying nonabelian gauge charge live in local 
representation spaces of the gauge group.
Gauge theories are covariant with respect to independent
basis changes (frame rotations) in the fibers over different
points of space--time. This covariance is achieved by introducing
a connection, or a gauge field,  which defines a parallel transport from
one fiber to another.

The mathematical framework of quantum (field) theory 
on the other hand is rather
different. Quantizing a free relativistic field theory, say, involves
selecting a foliation of space--like hypersurfaces in 
space--time, and an expansion of the field operators
in terms of the normal modes on these hypersurfaces.
The normal modes are in one--to--one correspondence with
the creation and annihilation operators which act on the
Hilbert space of states. It is quite obvious what ''nonlocality''
of the Hilbert space means in this case:
in the Schr\"odinger picture, its elements are wave functionals
whose arguments are functions defined over an entire space--like
hypersurface.

Inspired by the geometric structure of classical gauge theories
it is natural to ask if there exists a formulation of quantum
mechanics, or a generalization thereof, in which there is not only one
 global Hilbert space, but rather a bundle of Hilbert 
spaces with one such space associated to each point of the base manifold.
It is not clear a priori what this base manifold should be.
In a field theory context, a plausible choice is to identify
it with space--time. In particular
if one tries to construct a consistent theory of quantum gravity
it might prove helpful to recast the rules of quantum mechanics
in a language similar to that of classical general relativity.
There have been various suggestions along these lines in the
literature \cite{gr,ms}
but no complete theory has emerged so far. It seems
clear that using only classical field theory and standard 
quantum mechanics as an input does not lead to a unique theory,
and depending on which additional assumptions are made different
models with different physical interpretations arise.

Another option is associating a Hilbert space to each
point of {\it configuration--space}. A theory of this kind
could be applied to all quantum systems, not only to field
theories (in which case the configuration--space is infinite
dimensional). Such families of Hilbert spaces have played a certain
r\^ole in connection with Berry's phase \cite{berry} 
where the configuration--space
is the one pertaining to the slow degrees of freedom, but no
general theory has been developed so far.

In the present paper we investigate an even more general setting
where a local Hilbert space is ascribed to each point of 
{\it phase--space}. There are various motivations for this choice \cite{popov}.
The most important one is that this setting allows for an
intriguing 
reformulation of quantum mechanics which, while being strictly equivalent
to the usual one, gives a remarkable new interpretation  to the process
of ''quantization''.

The theory which we are going to develop is, on
the one hand, a Yang--Mills type gauge theory over phase--space
with an emphasis on local geometric structures. On the other
hand, it can be shown to be equivalent to standard quantum mechanics
with a single global Hilbert space. The typical fiber at each
point of phase--space is taken to be a copy of the ordinary 
quantum mechanical Hilbert space, henceforth denoted \cv.
In each one of those infinite dimensional spaces we can perform
independent changes of their bases by means of a unitary transformation $U$.
We shall denote local coordinates on phase--space by 
$\phi\equiv(\phi^a)$ and write $\cv_\phi$ for the Hilbert space at $\phi$.
Then the position--dependent unitary transformation $\phi\mapsto U(\phi)$
is precisely a local gauge transformation in the sense of Yang--Mills
theory. The gauge group is the infinite dimensional group of all
unitary transformations on \cv, and the corresponding Lie algebra
consists of hermitian operators satisfying the commutator relations
of a (generalized) $W_\infty$--algebra \cite{gelf,winf}. Hence a connection
can be locally represented by a 1--form
$\Gamma=\Gamma_a(\phi)d\phi^a$ where the ''gauge field''
$\Gamma_a(\phi)$ is a hermitian operator on $\cv_\phi$ 
(for $a$ and $\phi$ fixed).

The crucial question is which principle determines the
connection $\Gamma$. Is it dynamical as in the gauge 
theories which we use in particle physics, or is it
fixed to have a universal form? In this paper we shall demonstrate
that, to a large extent, $\Gamma$ is fixed by a deep physical
principle, invariance under the ''background--quantum split
symmetry''. We shall see that the implementation of this symmetry partially
replaces the usual process of quantization.

How can we reconcile then the standard single--Hilbert space
description of quantum mechanics with the picture of the local
Hilbert spaces drawn above? It is clear that if we had a parallel
transport at our disposal by means of which a vector in
$\cv_\phi$ can be transported consistently from $\phi$ to the
Hilbert space $\cv_{\bar\phi}$ at an arbitrary point $\bar\phi$, 
then the infinitely many Hilbert spaces were redundant. In this
case all vectors and operators of $\cv_{\bar\phi}$ can be obtained
from those of $\cv_\phi$ by a known unitary transformation.
This procedure is fully consistent only if the parallel transport
is path--independent, i.e., if the pertinent connection $\Gamma$
has a vanishing curvature. However, in our case it will not be
necessary to insist on a completely ''flat'' connection.
It is sufficient to require consistency
up to a physically irrelevant phase which means that the curvature
of $\Gamma$, $\Omega_{ab}(\Gamma)$, may be proportional to the unit
operator. Connections with this property are called {\it abelian}
since their curvature $\Omega_{ab}(\Gamma)$ commutes with any
other operator. We shall see that the connection which is 
dictated by the background--quantum split symmetry is indeed
an abelian one, and that the associated parallel transport can
be used in order to prove the equivalence of the local theory
with standard quantum mechanics.

In our approach the rules of canonical quantization are
replaced by two new, independent postulates. They are not
borrowed from standard quantum mechanics, but rather are
a mathematically very natural option if one works within the
gauge theory framework. This will shed  new light on
what is means to ''quantize'' a hamiltonian
system \cite{cons}. In a nutshell, our first postulate is
that in order to go from classical to quantum mechanics one
has to replace the vector representation of the group of linear
canonical transformations, $\Sp$, by its  spinor representation.
This leads us, by pure group theory, from classical mechanics to
the semiclassical approximation of quantum mechanics. The second
postulate is that the gauge theory should respect the 
background--quantum split symmetry. Imposing this symmetry, we
recover full--fledged, exact quantum mechanics from the semiclassical
theory resulting from the first postulate.
\vspace{3mm}

Let us be more explicit about
the ''matter fields'' which will appear in our gauge theory.
We choose them to be a generalization of the metaplectic
spinor fields, which are a kind of phase--space analogue
of the ordinary spinor fields over space--time. We consider
hamiltonian systems with $N$ degrees of freedom and a
$2N$--dimensional phase--space $\cm_{2N}$. Then tensor fields
over $\cm_{2N}$ transform under local frame rotations in the 
tangent spaces $T_\phi \cm_{2N}$ according to tensor products
of the vector representation of $\Sp$. Here the group $\Sp$
plays a r\^ole analogous to the Lorentz group
$SO(1,n-1)$, and the tensor fields on $\cm_{2N}$
are the analogs of the integer-spin fields on space--time.
Spinors on space--time, on the other hand, transform under
local frame rotations according to the double covering
of the Lorentz group, {\it Spin}$(1,n-1)$. Metaplectic spinors
on $\cm_{2N}$ are defined in a very similar fashion:
under local frame rotations they transform according to the
covering group of $\Sp$, i.e., the metaplectic group
$\Mp$. There exists a two--to--one homomorphism between
$\Mp$ and $\Sp$. Unlike {\it Spin}$(1,n-1)$ which has finite
dimensional matrix representations, the representations of
$\Mp$ are all infinite dimensional. We shall be interested
in unitary representations on the Hilbert space $\cv$.
The infinite dimensional space $\cv$ will serve as a typical
fiber for the Hilbert bundles we construct. At each point $\phi$
of $\cm_{2N}$ there will be a local tangent space 
$T_\phi\cm_{2N}$ and a local Hilbert space $\cv_\phi$ whose 
elements respond to a frame rotation at $\phi$ by a $\Sp$ and a
$\Mp$ transformation, respectively. 

In order to find representations of $\Mp$ we have to associate
to all matrices $S\equiv (S^a_{~b})\in \Sp$ a unitary
operator $M(S)$ such that $M(S_1)M(S_2)= \pm M(S_1 S_2)$ \cite{ko,lj,wood}.
These operators can be found by starting from the Clifford
algebra \cite{metino,superman}
\bel{1.1}
\gamma^a\gamma^b -\gamma^b\gamma^a=2i\,\omega^{ab}
\ee
where in terms of $N\times N$ blocks,
\renewcommand{\arraystretch}{1.3} 
\arraycolsep3pt
\bel{1.2}
(\omega^{ab})=\left(
\begin{array}{cc}0 & -1 \\ +1 & 0\end{array}
\right)
\ee
Here $\omega^{ab}$ is an antisymmetric analogue of the inverse metric
tensor on Minkowski space. The metaplectic gamma--''matrices'' $\gamma^a$
are supposed to be hermitian operators on $\cv$ which transform as a
vector of $\Sp$:
\bel{1.3}
M(S)^{-1}\gamma^aM(S)=S^a_{~b}\gamma^b
\ee
To solve this equation, we assume that $S$ is infinitesimally
close to the identity, i.e., that 
$S^a_{~b}=\delta^a_b+\omega^{ac}\kappa_{cb}$
with symmetric coefficients $\kappa_{ab}=\kappa_{ba}$. If we make the
ansatz
\bel{1.4}
M(S)=1-\tfrac{i}{2}\kappa_{ab}\Sigma^{ab}
\ee
then \rf{1.3} implies the following condition for the generators
$\Sigma^{ab}$:
\bel{1.5}
[\gamma^a , \Sigma^{bc}] = i (\omega^{ab}\gamma^c+\omega^{ac}\gamma^b)
\ee
It is easy check that this equation has the solution
\bel{1.6}
\Sigma^{ab}=\tfrac{1}{4}(\gamma^a\gamma^b+\gamma^b\gamma^a)
\ee
and that the generators satisfy the desired commutator relations:
\bel{1.7}
[\Sigma^{ab},\Sigma^{cd}]
=
i\left(
    \omega^{ac}\Sigma^{bd}+\omega^{bc}\Sigma^{ad}
   +\omega^{ad}\Sigma^{bc}+\omega^{bd}\Sigma^{ac}
  \right)
\ee
Thus every representation of the metaplectic Clifford algebra
in terms of hermitian $\gamma$--''matrices'' leads to hermitian
generators $\Sigma^{ab}=\Sigma^{ba}$ and to
unitary operators 
$\exp \left(-\tfrac{i}{2}\kappa_{ab}\Sigma^{ab}\right)
\in \Mp$ acting on \cv.

The crucial difference compared to spinors on Minkowski
space is the minus sign on the LHS of the Clifford algebra
\rf{1.1}. It has the consequence that there can be no finite
dimensional matrix representations, and it also means that
the metaplectic Clifford algebra is basically the same object as
the Heisenberg algebra. In fact, consider a set of position
operators $\hat{x}^k$ and momentum operators 
$\hat{\pi}^k$, $k=1,\cdots,N$, 
which act on $\cv$ and satisfy canonical commutation relations
$[\hat{x}^j,\hat{\pi}^k]=i\hbar \delta^{jk}$ with the other
commutators vanishing. If we combine $\hat{x}^k$ and $\hat{\pi}^k$
into
\bel{1.8}
\hat{\varphi}^a\equiv \left( \hat{\pi}^k, \hat{x}^k\right)
\quad , \quad a=1,\cdots,2N
\ee
then the commutation relations read
\bel{1.9}
[\hat{\varphi}^a, \hat{\varphi}^b] = i\hbar\, \omega^{ab}
\ee
Obviously we can realize the $\gamma$--''matrices'' in terms
of those position and momentum operators:
\bel{1.10}
\gamma^a=(2/\hbar)^{1/2}\,\hat{\varphi}^a
\ee
If we choose an arbitrary basis $\{|x\rangle \}$ in \cv,
$\gamma^a$ is represented by the matrix
\bel{1.11}
(\gamma^a)^x_{~y}=(2/\hbar)^{1/2}\, 
\langle x | \hat{\varphi}^a |y \rangle
\ee
We shall use both the matrix and the bra--ket notation, 
with bra (ket) vectors corresponding to upper (lower)
indices. The components of a vector $|\psi \rangle \in \cv$
are written as
\bel{1.12}
\psi^x = \langle x | \psi \rangle
\ee
and those of the dual vector 
$\langle \psi | \in \cv^*$ read
correspondingly
\bel{1.13}
\psi_x=\langle\psi | x\rangle =\left(\psi^x\right)^*
\ee
As the notation suggests already, we shall often
use the representation in which the $\hat{x}^k$'s are
diagonal and the label $x\equiv (x^k) \in \R^N$ is
the set of their eigenvalues. Then the dual pairing 
\bel{1.14}
\langle \chi | \psi \rangle \equiv \chi_x\psi^x
\equiv(\chi^x)^* \psi^x \equiv \int d^Nx\, \left(\chi^x\right)^*\psi^x
\ee
is the standard inner product on $L^2(\R^N, d^Nx)$.
It is convenient to use a formal matrix notation where the 
integration over repeated indices $x,y,\cdots$ is understood.
In this representation,
$\langle x | \hat{x}^k | y\rangle =x^k \delta(x-y) $
and 
$\langle x | \hat{\pi}^k | y\rangle =-i\hbar\partial_k\delta(x-y) $
so that the generators
\bel{1.15}
\Sigma^{ab}=\tfrac{1}{2\hbar}
\left(
\hat{\varphi}^a
\hat{\varphi}^b
+
\hat{\varphi}^b
\hat{\varphi}^a
\right)
\ee
become Schr\"odinger Hamiltonians (second order differential operators)
with a quadratic potential.

To summarize:
The tangent bundle $T\cm_{2N}$ has the base manifold $\cm_{2N}$,
and the fiber at the point  $\phi \in \cm_{2N}$, the tangent 
space $T_\phi \cm_{2N}$, is a copy of $\R^{2N}$. The structure group
$\Sp$ acts on this space in its vector representation. The associated
spin
bundle has the same base manifold, but the fiber at $\phi$ is the 
infinite dimensional space $\cv_\phi$, a copy of the Hilbert space \cv.
The structure group acts on $\cv_\phi$ in its spinor (i.e., metaplectic)
representation. The typical fiber $\cv$
can be realized as the space of square--integrable
functions $L^2(\R^N, d^Nx)$, for instance.

A section through the spin bundle is locally described by a
function
\bel{1.16}
\psi: \,\cm_{2N} \to \cv \quad , \quad \phi \mapsto |\psi\rangle_\phi
\in \cv_\phi
\ee
The notation $|\cdot \rangle_{\phi}$ indicates that the vector
$|\cdot \rangle_\phi$
is an element of the local Hilbert space at $\phi$. In a concrete
basis, this spinor field has the components
\bel{1.17}
\psi^x(\phi) = \langle x | \psi\rangle_\phi
\ee
If we use the dual space $\cv^*$ instead of $\cv$ we arrive at
the dual spin bundle. Locally its sections are functions
\bel{1.18}
\chi_x(\phi) = {}_\phi\la\chi | x \ra \quad , \quad 
{}_\phi\la \chi | \in \cv_\phi^*
\ee
We shall also consider multispinor fields
\bel{1.19}
\phi \mapsto \chi^{x_1\cdots x_r}_{y_1 \cdots y_s} (\phi)
\ee
which describe sections through the tensor product bundle
whose typical fiber is 
$\cv \otimes \cv \otimes \cdots \cv^* \otimes \cv^* \otimes \cdots$
with $r$ factors of $\cv$ and $s$ factors of $\cv^*$. We refer to 
\rf{1.19} as a $(r,s)$--multispinor. If we consider, for instance,
a family of operators which are labeled by the points
of $\cm_{2N}$, $A(\phi):\, \cv_\phi \to \cv_\phi$,
then its component representation
\bel{1.20}
A^x_{~y}(\phi) = \la x|A(\phi) | y\ra
\ee
is a $(1,1)$--multispinor field.

The operators $\hat{\varphi}^a=(\hat{\pi}^k, \hat{x}^k)$ should
not be confused with the momentum and position operators which
result from the canonical quantization of a system with 
phase--space $\cm_{2N}$. The $\hat\varphi$'s operate on the 
infinitely many fibers $\cv_\phi$ whose relation to the
quantum mechanical Hilbert space $\cv_{\rm QM}$
is not known yet. In the simplest case of a flat phase--space
$\cm_{2N}=\R^{2N}$ one can introduce canonical coordinates 
$\phi^a \equiv (p^k,q^k)$, and then quantization amounts
to introducing operators
$\hat{\phi}^a \equiv (\hat{p}^k, \hat{q}^k)$ with 
$[\hat{q}^j, \hat{p}^k]=i\hbar \delta^{jk}$.
The main issue of this paper will be to understand the
interrelation between $\hat{\phi}^a$ and $\hat\varphi^a$ 
and between $\cv_{\rm QM}$ and the family of spaces
$\{\cv_\phi, \, \phi \in \cm_{2N}\}$.

We shall see that conventional semiclassical quantization
provides a natural motivation for studying (generalized) metaplectic
spinors. It will turn out, however, that the transition
to full quantum mechanics makes it necessary to replace
the structure group $\Mp$ by the group of
{\it all } unitary transformations on \cv.

The remaining sections of this paper are organized 
as follows. In section 2 we show that switching from
the vector to the spinor representation brings us from
classical mechanics to the semiclassical quantum theory.
Then, in the following sections, we describe the construction
of full--fledged quantum mechanics by sewing  together
the local semiclassical approximations obtained in
this way. To do this, we need a connection on the bundle
of Hilbert spaces. It is introduced in section 3 where
also some of its properties are described which will 
be needed later on. In section 4 we demonstrate that,
within the present approach, exact quantum mechanics
is recovered from the semiclassical approximations
if one requires invariance
under the background--quantum split symmetry. This condition
forces the connection to be abelian. We show that by virtue
of the corresponding parallel transport all local Hilbert 
spaces can be identified with a single reference Hilbert
space, and that the resulting reduced theory coincides
with conventional quantum mechanics. In section 5 we
continue the discussion of the bundle approach for the
special case of a flat phase--space, and section 6 contains
the conclusions.

For a first reading, the reader who is mostly interested
in the general results can proceed directly to the
summary of section 6 before studying the details of
their derivation.

\mysection{Lie--Derivative and Semiclassical    
Approximation }
In this section we construct a Lie--derivative for 
metaplectic spinor fields with respect to an arbitrary
globally hamiltonian vector field. This Lie--derivative
is the natural link between the Hilbert bundle approach
and the semiclassical approximation of standard quantum mechanics.
From a slightly different point of view, it has been discussed
in ref.\cite{metino, metone} already.

We work on an arbitrary symplectic manifold ($\cm_{2N}, \omega$).
By definition \cite{mar} the symplectic 2--form
\bel{2.1}
\omega =\tfrac{1}{2}\omega_{ab}(\phi) d\phi^a\wedge d\phi^b
\ee
is closed, $d\omega=0$, and nondegenerate, i.e., the matrix
$(\omega_{ab})$ possesses an inverse $(\omega^{ab})$:
\bel{2.2}
\omega_{ab}\omega^{bc}=\delta^c_a
\ee
We may choose Darboux local coordinates
$\phi^a=(p^k,q^k)$, $k=1,\cdots,N$ with respect to which
$\omega_{ab}$ is constant:
\bel{2.3}
(\omega_{ab})=
\left(
\begin{array}{cc}
0 & +1 \\
-1 & 0
\end{array}
\right)
\ee
The inverse of this matrix is precisely \rf{1.2} which
appeared in the Clifford algebra. A globally hamiltonian
vector field $h=h^a\partial_a$, $\partial_a\equiv\partial/\partial\phi^a$,
can be expressed in terms of a generating function $H:\cm_{2N}\to \R$
according to 
\bel{2.4}
h^a=\omega^{ab}\partial_b H
\ee
The flow generated by $h$ leaves the form $\omega$ invariant,
i.e., the (ordinary) Lie--derivative of $\omega$ with respect
to $h$ vanishes:
\bel{2.5}
\ell_h\omega_{ab}\equiv h^c \,\partial_c \omega_{ab}
                       +\partial_ah^c\, \omega_{cb}
                       +\partial_bh^c\, \omega_{ac}=0
\ee

Quite generally, the Lie--derivative for any tensor
field $\chi$ on an arbitrary $n$--dimensional manifold
$\cm_n$ is of the form \cite{dewitt}
\bel{2.6}
\ell_v \chi = v^a\partial_a \chi -\partial_b v^a\,G^b_{~a}\chi
\ee
Here $v=v^a\partial_a$ is an arbitrary vector field,
and the matrices $G^b_{~a}$ are the generators of
$GL(n, \R)$ in the representation to which $\chi$ belongs.
They obey the Lie algebra relations
\bel{2.7}
[G^b_{~a} , G^c_{~d}] =\delta^b_d\, G^c_{~a}- \delta^c_a\, G^b_{~d}
\ee
Lie--derivatives form a closed algebra under commutation:
\bel{2.8}
[\ell_{v_1}, \ell_{v_2}] =\ell_{[v_1,v_2]}
\ee
Here
\bel{2.9}
[v_1,v_2]^a \equiv v^b_1\, \partial_b v^a_2 - v_2^b\,\partial_b v_1^a
\ee
denotes the Lie--bracket. The Lie--derivative is a covariant
differentiation which needs no connection for its
definition; if $\chi$ transforms as a tensor under general
coordinate transformation, so does $\ell_v\chi$.

For spinors the situation is more complicated. Under
general coordinate transformations they behave like
scalars, $\delta_C \psi =v^a\partial_a\psi$, but they
transform nontrivially under local frame rotations:
$\delta_F\psi=-\frac{i}{2}\kappa_{ab}\Sigma^{ab}\psi$.
(The generators $\Sigma^{ab}$ and parameters $\kappa_{ab}$
are antisymmetric in $a$ and $b$ for $O(n$) and symmetric for
$\Sp$.)
One can try to construct a Lie--derivative for $\psi$ by
combining a general coordinate transformation with an 
appropriate frame rotation, and to express the 
$G^a_{~b}$'s in eq.\rf{2.6} in terms of the generators
$\Sigma^{ab}$.
In general this is possible only for a restricted class
of vector fields.

On {\it Riemannian manifolds} the construction can be 
performed only if $v$ is a Killing vector field. In this
case $\Sigma^{ab}$ and $G^b_{~a}$ are related by a contraction
with the metric, which  has a vanishing Lie--derivative.
As a consequence, the resulting $\ell_v$ has all the formal
properties of a Lie--derivative \cite{dewitt}.

On {\it symplectic manifolds} a spinorial Lie--derivative can
be defined for a much larger (infinite dimensional, in fact)
space of vector fields, namely for all globally hamiltonian
vector fields. For them, $\ell_v\omega=0$, and we may use
$\omega_{ab}$ and $\omega^{ab}$ in order to relate $G^b_{~a}$
to $\Sigma^{ab}$. This is particularly
clear in Darboux coordinates. If we set
\bel{2.10}
\Sigma^{ab}=i\left(
G^a_{~c} \,\omega^{cb} + G^b_{~c}\,\omega^{ca}
\right)
\ee
it can be checked that the algebras \rf{2.7} and \rf{1.7}
for $G^a_{~b}$ and $ \Sigma^{ab}$, respectively, become
equivalent. Now we specialize
\bel{2.11}
\ell_h=h^a\partial_a -\partial_b h^a \,G^b_{~a}
\ee
for the hamiltonian vector field \rf{2.4} and use \rf{2.10}:
\bel{2.12}
\ell_H \equiv \ell_h =h^a(\phi) \partial_a
                     +\tfrac{i}{2}\partial_a\partial_b H(\phi)
                      \,\Sigma^{ab}
\ee
This equation defines a Lie--derivative for fields transforming
in an arbitrary representation of the algebra \rf{1.7},
for spinors in particular.
Usually we deal with tensor products of the vector and the
spinor representation and their respective duals.
Then we have explicitly
\bel{2.13}
\bay
\ell_H 
\chi^{a\cdots x\cdots}_{b\cdots y \cdots}
&=&\dis
 h^c\partial_c \chi^{a\cdots x\cdots}_{b\cdots y \cdots}
-\partial_c h^a\chi^{c\cdots x\cdots}_{b\cdots y \cdots}
+\partial_b h^c \chi^{a\cdots x\cdots}_{c\cdots y \cdots}
\\[.3cm]
&&\dis
+\tfrac{i}{2}\partial_c\partial_d H\left(\Sigma^{cd}\right)^x_{~z}
 \chi^{a\cdots z\cdots}_{b\cdots y \cdots}
-\tfrac{i}{2}\partial_c\partial_d H
 \chi^{a\cdots x\cdots}_{b\cdots z \cdots}
\left(\Sigma^{cd}\right)^z_{~y}
+\cdots
\eay
\ee
Here $\Sigma^{ab}$ has now the more concrete meaning of the
generators \rf{1.6} written in terms of the 
$\gamma$--matrices \rf{1.10}.
It is easy to check that \rf{2.13} defines a covariant
derivation. If we replace all partial derivatives in $\ell_H$
by covariant derivatives with respect to a symplectic 
connection (see below) all terms involving the connection
are seen to cancel. Moreover, as a consequence of \rf{2.8},
the $\ell_H$'s form a representation of the infinite--dimensional
Lie algebra of symplectic diffeomorphisms:
\bel{2.14}
\left[ \ell_{H_1} , \ell_{H_2}\right] = 
-\ell_{\{H_1, H_2\} }
\ee
Here
\bel{2.15}
\left\{H_1, H_2 \right\} =\partial_a H_1 \omega^{ab} \partial_b H_2
\ee
denotes the Poisson bracket.

Sometimes the bra--ket notation is more convenient than the 
component notation. For $\psi^x = \la x| \psi \ra_\phi$, say,
\bel{2.16}
\ell_H |\psi\ra_\phi =h^a \partial_a |\psi\ra_\phi
+\tfrac{i}{2}\partial_a\partial_b H\, \Sigma^{ab}|\psi\ra_\phi
\ee
The Lie--derivative of the operator field \rf{1.20} reads likewise
\bel{2.17}
\ell_H A(\phi) = h^a\partial_a A(\phi) 
+\tfrac{i}{2}\partial_a\partial_b H \left[\Sigma^{ab}, A(\phi)\right]
\ee
Applying \rf{2.13} to $\gamma^{ax}_{~~y}$ we find that the
$\gamma$--matrices and the $\hat\varphi^a$'s have vanishing
Lie--derivatives:
\bel{2.18}
\ell_H \gamma^a =0\quad , \quad \ell_H \hat\varphi^a=0
\ee

Let us now turn to the physical interpretation of $\ell_H$.
Given the vector field $h^a$, we can introduce a one--particle
dynamics on $\cm_{2N}$ by virtue of Hamilton's equation of 
motion (the dot denotes the time derivative):
\bel{2.19}
\dot\phi^a(t)=h^a\left(\phi(t)\right)
\ee
The equivalent many--particle description is provided by 
Liouville's
equation $\partial_t \rho =\{ H,\rho \}= - h^a \partial_a \rho$
for the probability density $\rho\equiv\rho(\phi;t)$.
In standard symplectic geometry one generalizes Liouville's 
equation to arbitrary tensor fields $\chi$,
whose evolution along the hamiltonian flow involves
the ordinary Lie--derivative: $\partial_t \chi =-\ell_h \chi$.
It is therefore very natural to investigate the analogous
time--evolution for more general fields 
$\chi\equiv \left(\chi^{a\cdots x \cdots}_{b\cdots y \cdots}\right)$,
\bel{2.20}
-\partial_t\chi(\phi;t)=\ell_H \chi (\phi;t)
\ee
with $\ell_H$ given by 
\rf{2.13}\footnote{
See refs.\cite{cpi} and \cite{metino} for a path--integral
solution of eq.\rf{2.20} in the case of $p$--forms
and of antisymmetric $(0,s)$--multispinors, respectively.
The latter path--integral describes a kind of topological field
theory which was used \cite{metone} to detect obstructions
for the existence of metaplectic spin structures on 
$\cm_{2N}$ \cite{spinstru,witten}.
}.
For example,
\bel{2.21}
-\partial_t \psi^x(\phi;t)=h^a\partial_a\psi^x(\phi;t)
+\tfrac{i}{2}\partial_a\partial_b H(\phi)
\left(\Sigma^{ab}\right)^x_{~y} \psi^y(\phi;t)
\ee

Up to now we considered metaplectic spinors which are
defined at each point $\phi$ of $\cm_{2N}$. Equally
important are spinors which are defined only along
a certain curve in phase--space.
In a slight abuse of language we shall refer to them
as ''world--line spinors''. We are particularly
interested in the case where the curve is given by
some solution $\Phi_{\rm cl}^a(t)$ of Hamilton's
equation \rf{2.19}. Let $\eta(t)$ be 
a world--line spinor along this classical trajectory, i.e.,
for different times $\eta(t)$ lives in different Hilbert spaces:
$$\eta(t)\in \cv_{\Phi_{\rm cl}(t)}.$$
Given a world--line spinor, we can define a singular spinor
{\it field} by writing
\bel{2.22}
\psi^x(\phi ; t) =\eta^x(t)\, \delta\left(\phi -\Phi_{\rm cl}(t)\right)
\ee
We demand that $\psi^x$ satisfies the evolution equation
\rf{2.21}. This entails the following equation of motion
for $\eta^x$ and its dual $\bar\eta_x=(\eta^x)^*$:
\bel{2.23}
\bay
\partial_t\eta^x(t) & = & \dis 
-\tfrac{i}{2} \partial_a \partial_b H\left( \Phi_{\rm cl}(t)\right)
\left(\Sigma^{ab}\right)^x_{~y} \eta^y(t) \\
\partial_t{\bar\eta}_x(t)& = &\dis
\;\; \tfrac{i}{2} \partial_a \partial_b H
\left( \Phi_{\rm cl}(t)\right)
\bar\eta_y(t) 
\left(\Sigma^{ab}\right)^y_{~x} 
\eay
\ee
There exists an interesting relation between \rf{2.23}
and the corresponding equation of motion for a world--line
vector field $c^a(t)$:
\bel{2.24}
\partial_t c^a(t) =\partial_b h^a \left( \Phi_{\rm cl}(t)\right) c^b(t)
\ee
In this case
$ c(t) \in T_{\Phi_{\rm cl}(t)} \cm_{2N}$
lives in the tangent spaces along the classical path. With the
ansatz
\bel{2.25}
V^a(\phi;t) =c^a(t)\, \delta\left(\phi- \Phi_{\rm cl}(t)\right)
\ee
eq.\rf{2.24} is indeed equivalent to $-\partial_tV^a=\ell_hV^a$.
Eq.\rf{2.24} is precisely Jacobi's equation which governs
small {\it classical} fluctuations about the trajectory
$\Phi_{\rm cl}(t)$. In fact, if we write 
$\Phi_{\rm cl}^\prime(t)=\Phi_{\rm cl}(t)+c(t)$
and require that, to first order in $c$, also $\Phi^\prime_{\rm cl}$
solves Hamilton's equation, then the ''Jacobi field'' $c(t)$
must obey \rf{2.24}. The world--line spinors have the 
remarkable property of being a kind of ''square root'' of
the Jacobi--fields. If $\eta$ and $\bar\eta$ transform in the 
spinor representation of $\Sp$ and its dual, respectively, it
is clear that 
$\bar\eta \gamma^a \eta \equiv (2/\hbar)^{1/2}\bar\eta\hat\varphi^a\eta$
transforms as a vector. Furthermore, if one sets
\bel{2.26}
c^a(t) = \bar\eta(t) \,\hat\varphi^a\,\eta(t)
\ee
and uses the equation of motion for $\eta$ and $\bar\eta$,
eq.\rf{2.23}, then it follows that \rf{2.26} is a solution of
Jacobi's equation \rf{2.24}. This fact finds a natural 
interpretation in the context of the semiclassical quantization
which we discuss next.

We consider a quantum system with phase--space $\cm_{2N}$
and classical Hamiltonian 
$H(\phi^a) =H(p^k ,q^k)$\footnote{For simplicity we assume
here that $\cm_{2N}=\R^{2N}$ is the symplectic plane 
which can be covered by a single
chart of Darboux coordinates $\phi^a\equiv(p^k,q^k)$,
$a=1\cdots 2N$, $k=1\cdots N$.}.
The probability amplitude for a transition between two points
in configuration space, $q_1^k$ and $q_2^k$, is given by the
path--integral
\bel{2.27}
\la q_2,t_2 | q_1,t_1 \ra_H
=
\int \cd  p(t) \int \cd q(t) 
\exp\left[
\frac{i}{\hbar}\int_{t_1}^{t_2} dt\,
\left\{p^k\dot{q}^k -H(p,q)\right\}
\right]
\ee
subject to the boundary conditions $q(t_{1,2})=q_{1,2}$.
Let us shift the variable of integration 
$\phi(t)\equiv \left(p(t), q(t)\right)$
by an arbitrary solution of Hamilton's equation,
$\Phi_{\rm cl}(t)\equiv \left(p_{\rm cl}(t), q_{\rm cl}(t)\right)$:
\bel{2.28}
\phi^a(t)=\Phi_{\rm cl}^a(t)+\varphi^a(t)
\quad , \quad
\varphi^a(t)\equiv \left(\pi^k(t), x^k(t)\right)
\ee
Inserting this shift on the RHS of \rf{2.27} and
using $\dot\Phi_{\rm cl}^a = h^a(\Phi_{\rm cl})$ we 
obtain

\bel{2.29}
\la q_2 , t_2 | q_1 ,t_1\ra_H
=
\exp\left[ \tfrac{i}{\hbar}
\left\{ S_\rcl +p^k_\rcl (t_2) x^k_2
-p^k_\rcl(t_1) x_1^k \right\} 
\right]
 \la x_2,t_2 |x_1,t_1
\ra_{\cal H}
\ee
Here 
$S_\rcl \equiv \int dt\,\left\{p^k_\rcl \dot{q}^k_\rcl
-H(p_\rcl,q_\rcl)\right\}$
is the action along the classical trajectory; furthermore
we defined the shifted path--integral
\bel{2.30}
\la x_2, t_2 | x_1, t_1\ra_{\cal H}
\equiv \int \cd\pi(t)
\!\!\!\!\!\!\!\!\!
\int\limits_{x(t_{1,2})=x_{1,2}}
\!\!\!\!\!\!\!\!\!
\cd x(t) 
\exp \left[
 \frac{i}{\hbar}\int_{t_1}^{t_2} dt\,
\left\{\pi^k\dot{x}^k-\ch(\varphi^a; \Phi_\rcl^a)\right\}
\right]
\ee
with its Hamiltonian
\bel{2.31}
\ch(\varphi; \Phi_\rcl)
\equiv H(\Phi_\rcl+\varphi)-\varphi^a\partial_aH(\Phi_\rcl)
-H(\Phi_\rcl)
\ee
and the boundary values
\bel{2.32}
x_{1,2}\equiv q_{1,2}-q_\rcl(t_{1,2})
\ee
Note that the shifted path--integral \rf{2.30} is a
solution of the Schr\"odinger 
equation\footnote{Here and in the following we assume that
all operators are Weyl--ordered and that the 
path--integrals are discretized correspondingly
(mid--point rule).}
\bel{2.33}
\left[
i\hbar \,\partial_t -\ch\left(\hat\varphi^a ; \Phi^a_\rcl\right)\right]
\la x,t | x_1 ,t_1
\ra_{\cal H}=0
\ee
The operators $\hat\varphi^a\equiv(\hat\pi^k, \hat{x}^k)$
result from the ordinary canonical quantization of the
$\varphi^a$--degrees of freedom; they satisfy the canonical
commutation relations
$\left[ \hat\varphi^a, \hat\varphi^b\right]=i\hbar \omega^{ab}$.
In eq.\rf{2.33} their representation in terms of multiplication
and derivative operators is employed:
$\hat{x}^k=x^k$, $\hat\pi^k=-i\hbar\partial/\partial x^k$.

Up to this point, no approximation has been made. In a traditional
semiclassical calculation one would expand about a classical
trajectory connecting $q_1$ and $q_2$ for which $x_1=x_2=0$ therefore,
and one would expand $\ch(\varphi;\Phi_\rcl)$ with respect to the
fluctuation $\varphi$; to lowest order, only the quadratic term
is kept:
\bel{2.33a}
\ch\left(\varphi; \Phi_\rcl\right)
=\tfrac{1}{2} \partial_a\partial_b 
H\!\left(\Phi_\rcl\right)\varphi^a\varphi^b+O(\varphi^3)
\ee
Here we shall not assume that $x_1$ and $x_2$ are zero
exactly but only that the transition amplitude is dominated
by a classical trajectory which passes near $q_1$ and $q_2$
at $t=t_1$ and $t=t_2$, respectively.

It is an important observation that the operator version
of the approximated Hamiltonian \rf{2.34} lies in the 
Lie algebra of $\Mp$. In fact, the operators
$\hat\varphi^a$ which appear naturally in the operatorial
formulation of the quantum mechanics defined by the
shifted path--integral \rf{2.30} can be identified with
the $\hat\varphi$'s which were introduced in section 1
as a realization of the metaplectic $\gamma$--matrices.
They are related to the generators of $\Mp$ by
eq.\rf{1.15}. Hence the Hamiltonian operator which governs
the $\varphi$--dynamics in the quadratic approximation
reads
\bel{2.34}
\bay
\ch\left(\hat\varphi ; \Phi_\rcl \right)
&=& \dis
\tfrac{1}{2}\,\partial_a\partial_b H(\Phi_\rcl)\, 
\hat\varphi^a\hat\varphi^b +\cdots
\\
&=& \dis
\tfrac{\hbar}{2}\,\partial_a\partial_b H(\Phi_\rcl) 
\,\Sigma^{ab}+\cdots
\eay
\ee

Given the matrix elements
$\la x,t| x_1, t_1\ra_{\cal H}$
we can fix $x_1$ and $t_1$ and define a world--line
spinor along $\Phi_\rcl (t)$ by
\bel{2.35}
\eta^x(t)\equiv \la x| \eta(t)\ra_{\Phi_\rcl(t)}
\equiv \la x,t| x_1 ,t_1\ra_{\cal H}
\ee
For every fixed time $t$, $\eta(t)$ lives in the fiber
$\cv_{\Phi_\rcl(t)}$. The justification for calling the
transition matrix element a world--line spinor is that 
with the semiclassical Hamiltonian \rf{2.34} the
Schr\"odinger equation \rf{2.33} is exactly the same
as the original equation of motion of world--line
spinors, eq.\rf{2.23}.

To summarize: For globally hamiltonian vector fields the
notion of a Lie--derivative and the corresponding 
transport along the hamiltonian flow can be generalized
from tensor to metaplectic spinor fields. For spinor
fields which have support only along classical trajectories,
this transport induces a well--defined equation of motion
for the ''world--line'' spinors. Semiclassical wave functions,
describing quantum fluctuations about classical trajectories, 
are found to be world--line spinors in this sense.

Thus we see that semiclassical quantum mechanics is 
most naturally formulated in terms of a family of
Hilbert spaces along the classical trajectory, $\cv_{\Phi_\rcl(t)}$.
In the full quantum theory we integrate over all paths
$\phi(t)$, and generically classical trajectories do not play
any preferred r\^ole. Therefore it is plausible to
conjecture that the generalization of the above picture to 
full--fledged quantum mechanics will involve a family of
Hilbert spaces $\cv_\phi$, $\phi\in \cm_{2N}$,  with a 
copy of $\cv$ at {\it all} points of phase--space. Before we
can investigate this question we have to introduce the 
notion of a spin connection.
\mysection{Spin Connections}
\subsection{General hermitian spin connection}
We saw that sections through the Hilbert bundle are locally
given by functions $\psi^x(\phi)=\la x | \psi \ra_\phi$.
Let us find a covariant derivative of the form
\bel{3.1}
\nabla_a|\psi\ra_\phi =\partial_a |\psi\ra_\phi
+\tfrac{i}{\hbar}\Gamma_a(\phi) |\psi \ra_\phi
\ee
For the dual spinor we set 
\bel{3.2}
\nabla_a\,{}_\phi\la\psi| = (\nabla_a|\psi\ra_\phi)^\da
\ee
where the adjoint is with respect to the inner product
of $\cv_\phi$. A priori, $\{\Gamma_a(\phi),\; a=1,\cdots, 2N \}$
is a set of $2N$
arbitrary operators on $\cv_\phi$. We require
that for all $|\psi \ra_\phi \in \cv_\phi$ and 
${}_\phi\la\chi |\in\cv_\phi^*$
\bel{3.3}
\nabla_a \,{}_\phi \la \chi | \psi\ra_\phi =\partial_a\,
{}_\phi\la \chi |\psi\ra_\phi
\ee
Then eqs.\rf{3.1}, \rf{3.2} and the Leibniz rule for $\nabla_a$
imply that the $\Gamma_a$'s must be hermitian:
\bel{3.4}
\Gamma_a(\phi) =\Gamma_a(\phi)^\da
\ee
For the time being we do not impose any further conditions on
$\Gamma_a$. Hence the Lie algebra of the structure group
(gauge group) is defined to be the space $\cg$ of all 
hermitian operators on \cv.
The infinite dimensional gauge group $G$ is the group of 
all unitary operators on \cv. It has $\Mp$ as a finite
dimensional subgroup. By using \cg--valued connections
we generalize the idea of ''local frame rotations'' in a 
way appropriate for Hilbert spaces; here ''frame'' means
a basis of \cv. In all fibers $\cv_\phi$ we may perform
independent changes of their bases by means of a 
gauge transformation
$U:\cm_{2N} \to G$, $\phi\mapsto U(\phi)$. It acts
according to
\bel{3.5}
|\psi\ra_\phi^\prime =U(\phi) |\psi\ra_\phi
\ee
From the covariance of $\nabla_a \equiv \nabla_a(\Gamma)$,
\bel{3.6}
\nabla_a(\Gamma^\prime) =U(\phi) \,\nabla(\Gamma)\, U(\phi)^{-1}
\ee
we obtain the transformation law of $\Gamma_a$:
\bel{3.7}
\Gamma_a^\prime(\phi)= U(\phi)\, \Gamma_a(\phi)\, U(\phi)^{-1}
+i\hbar\,\partial_a \,U(\phi)\, U(\phi)^{-1}
\ee
For an infinitesimal gauge transformation 
$U(\phi)=1-i\varepsilon(\phi)/\hbar$, $\varepsilon=\varepsilon^\da$,
one has
\bel{3.8}
\bay
\delta_F |\psi\ra_\phi& = &\dis -\tfrac{i}{\hbar}\,\varepsilon\,
 |\psi \ra_\phi
\\
\delta_F \Gamma_a(\phi)& =& \dis \partial_a \varepsilon 
+\tfrac{i}{\hbar} \left[ \Gamma_a , \varepsilon \right]
\eay
\ee
The local frame rotations are Yang--Mills-type gauge transformations
which are unrelated to coordinate transformations on $\cm_{2N}$.
Under a coordinate transformation $\delta_C\phi^a=-h^a(\phi)$
the spinor transforms as a scalar and $\Gamma_a$ as a vector:
\bel{3.8a}
\bay
\delta_C |\psi\ra_\phi& = &\dis h^a\,\partial_a |\psi \ra_\phi
\\
\delta_C \Gamma_a(\phi)& =&\dis h^b\,\partial_b \Gamma_a
+\partial_a h^b \Gamma_b
\eay
\ee

If we write the components of $\Gamma_a$ with respect to
an arbitrary basis $\{ |x\ra \}$ of $\cv$ in the form
\bel{3.9}
\Gamma_a(\phi)^x_{~y} = \la x |\Gamma_a(\phi) | y \ra
\ee
then eq.\rf{3.1} becomes
\bel{3.10}
\nabla_a \psi^x(\phi) =\partial_a \psi^x(\phi) 
+\tfrac{i}{\hbar}\, \Gamma_a(\phi)^x_{~y}\, \psi^y(\phi)
\ee
More generally we can consider $(r,s)$ multispinors 
$\chi^{x_1\cdots x_r}_{y_1\cdots y_s}$
which respond to a gauge transformation as
\bel{3.11}
\chi^{x\cdots }_{y\cdots }(\phi)^\prime
=U(\phi)^x_{~v} \,U^\da(\phi)^w_{~y}\, \cdots\,
\chi^{v\cdots }_{w\cdots }(\phi)
\ee
Their covariant derivative reads
\bel{3.12}
\nabla_a\chi^{x\cdots }_{y\cdots }
=
\partial_a\chi^{x\cdots }_{y\cdots }
+\tfrac{i}{\hbar}\,\Gamma_a(\phi)^x_{~z}\,\chi^{z\cdots }_{y\cdots }
-\tfrac{i}{\hbar}\,
\chi^{x\cdots }_{z\cdots }\,
\Gamma_a(\phi)^z_{~y}\,
+\cdots
\ee
For an $(1,1)$--multispinor $A(\phi)^x_{~y}$, interpreted
as an operator on $\cv_\phi$, one has for instance
\bel{3.13}
\nabla_a A =\partial_a A+\tfrac{i}{\hbar}\left[\Gamma_a, A \right]
\ee
Hence the second  equation of \rf{3.8} is simply
$\delta_F \Gamma_a =\nabla_a\varepsilon$.

Furthermore, $\Gamma_a$ gives rise to an exterior derivative
$\nabla=d\phi^a\nabla_a$ of $(r,s)$--spinor--valued 
differential forms
\bel{3.14}
F=F_{a_1\cdots a_p}(\phi)\, d\phi^{a_1}\cdots d\phi^{a_p}
\ee
which is covariant with respect to both Yang--Mills gauge
transformations and diffeomorphisms of $\cm_{2N}$.
The components $F_{a_1\cdots a_p}$ are multispinor fields.
We define
\bel{3.15}
\nabla F^{x\cdots}_{y\cdots}
=
\nabla F^{x\cdots}_{y\cdots,\, a_1\cdots a_p}(\phi)
\,d\phi^{a_1}\cdots d\phi^{a_p}
\ee
with the derivative of the components given by \rf{3.12}.
For an operator--valued $p$--form, say,
\bel{3.16}
\nabla A =dA+\tfrac{i}{\hbar} \,[\Gamma, A]
\ee
Here $ d\equiv d\phi^a\partial_a$ is the usual exterior derivative
and
\bel{3.17}
\Gamma\equiv \Gamma_a(\phi) d\phi^a
\ee
denotes the connection 1--form. In equations such as
\rf{3.16} the square brackets denote the graded commutator
\bel{3.18}
[A,B]=AB-(-1)^{[A][B]} BA
\ee
with $[A]=+1$ $([A]=-1)$ if the rank of $A$ as a differential
form is even (odd). The product $AB$ is a combination of
operator multiplication and wedge product.
For any two operator--valued differential forms
$A$ and $B$ one has the product rule
\bel{3.19}
\nabla(AB)= (\nabla A)B+(-1)^{[A]}A\nabla B
\ee

The curvature of $\Gamma_a$ is given by the hermitian
operators
$\Omega_{ab}=-i\hbar[\nabla_a, \nabla_b]$,
or explicitly,
\bel{3.21}
\Omega_{ab}=\partial_a\Gamma_b-\partial_b\Gamma_a
+\tfrac{i}{\hbar}\left[\Gamma_a, \Gamma_b \right]
\ee
with the components
\bel{3.22}
\Omega_{ab~y}^{~~x}
=
\partial_a\Gamma_{b~y}^{~x} 
-\partial_b\Gamma_{a~y}^{~x}
+\tfrac{i}{\hbar}
\left(
\Gamma_{a~z}^{~x}
\Gamma_{b~y}^{~z}
-
\Gamma_{b~z}^{~x}\Gamma_{a~y}^{~z} 
\right)
\ee
The curvature 2--form
\bel{3.23}
\Omega=\tfrac{1}{2}\Omega_{ab}(\phi)\, d\phi^a d\phi^b
\ee
reads in terms of $\Gamma$:
\bel{3.24}
\Omega=d\Gamma+\tfrac{i}{\hbar}\,\Gamma^2
\ee
Using this representation together with \rf{3.16}
it is easy to show that $\Omega$ satisfies Bianchi's identity
\bel{3.25}
\nabla\Omega \equiv d\Omega+\tfrac{i}{\hbar}\, [\Gamma, \Omega] =0
\ee
and that for any operator--valued $p$--form $A$
\bel{3.26}
\nabla^2A=\tfrac{i}{\hbar}\, [\Omega, A]
\ee
Under the gauge transformation \rf{3.5},
\bel{3.27}
\Omega_{ab}^\prime(\phi) = U(\phi)\, \Omega_{ab}(\phi)\, U(\phi)^\da
\ee 
\subsection{Symplectic connections}
Next we introduce a special class of spin connections
which assume values in the Lie algebra of the most important
subgroup of $G$, namely $\Mp$.

While in most parts of this paper Darboux local coordinates
are used, we shall employ a generic system of local coordinates
$\phi^a$ in this subsection. Hence $\omega_{ab}(\phi)$ will not
be given by the constant canonical matrix \rf{2.3} in general.
For clarity, the latter is denoted
\bel{3.28}
\left(\omcirc_{\alpha\beta}\right)=
\left(
\begin{array}{cc} 
0 & +1 \\
-1 & 0
\eay
\right)
\ee
here. We introduce a basis of vielbein fields 
$e_\alpha=e_\alpha^a\partial_a$ and of dual
1--forms $e^\alpha=e^\alpha_ad\phi^a$, $\alpha=1,\cdots,2N$.
Their components satisfy
\bel{3.29}
e^a_\alpha \,e^\beta_a=\delta^\beta_\alpha
\qquad , \qquad e^a_\alpha \,e^\alpha_b =\delta^a_b
\ee
In this subsection we make a notational distinction
between coordinate indices $a,b,c,\cdots$ and frame
indices $\alpha,\beta,\gamma,\cdots$.
We identify the vielbeins with the transformation which
brings $\omega_{ab}(\phi)$ to the skew--diagonal form
$\omcirc_{\alpha\beta}$:
\bel{3.30}
\omega_{ab}(\phi)=e^\alpha_a(\phi) e^\beta_b(\phi) 
\omcirc_{\alpha\beta}
\ee
Clearly $\omega_{ab}$ and $\omcirc_{\alpha\beta}$ are 
analogous to the metric on a curved space and on flat
space, respectively.
The group $\Sp$ acts on the frame indices
with matrices $S^\alpha_{~\beta}$ preserving 
$\omcirc$: 
$\omcirc_{\alpha_\beta}S^\alpha_{~\gamma}S^\beta_{~\delta}
=\omcirc_{\gamma\delta}$.
An infinitesimal local $\Sp$--rotation reads
\bel{3.31}
S^\alpha_{~\beta}(\phi) =\delta^\alpha_\beta+
\omcirc^{\alpha\gamma}\kappa_{\gamma\beta}(\phi)
\ee
with symmetric parameters $\kappa_{\alpha\beta}=\kappa_{\beta\alpha}$.
(Recall that the matrices in $\Sp$ are products of 
$\omcirc^{\alpha\beta}$ with some symmetric matrix \cite{lj}.
If $G^\alpha_{~\beta}$ is any generator of $\Sp$
in its defining representation, than $\omcirc_{\alpha\gamma}G^\gamma_{~\beta}$
is symmetric in $\alpha$ and $\beta$.)

Let us introduce a symplectic connection on the tensor bundle
over $\cm_{2N}$. By definition \cite{fedbook},
this is a torsion--free connection $\Gamma_{ab}^{c}=\Gamma_{ba}^{c}$
which preserves the symplectic structure:
\bel{3.32}
\nabla_c\,\omega_{ab} =\partial_c\omega_{ab} 
-\Gamma_{ca}^{d}\,\omega_{db}-\Gamma_{cb}^{d}\,\omega_{ad}=0
\ee
We introduce a related connection $\Gamma_{a~\beta}^{~\alpha}$
on the frame bundle in such a way that the (total) covariant
derivative of the vielbein vanishes:
\bel{3.33}
\nabla_a \,e^\alpha_b =\partial_a e^\alpha_b
-\Gamma_{ab}^{c}\, e^\alpha_c 
+\Gamma_{a~\beta}^{~\alpha} \,e^\beta_b=0
\ee
This fixes $\Gamma_{a~\beta}^{~\alpha}$ in terms of
$\Gamma_{ac}^{b}$:
\bel{3.34}
\Gamma_{a~\beta}^{~\alpha}=e^\alpha_c\,
\Gamma_{ab}^{c}\, e^b_\beta -e^b_\beta\, \partial_a e^\alpha_b
\ee
Since both $\omega_{ab}$ and $e^\alpha_a$ have vanishing covariant
derivative, eq.\rf{3.30} implies that
\bel{3.35}
\nabla_a\omcirc_{\alpha\beta}=
-\left(\omcirc_{\gamma\beta}\,\Gamma_{a~\alpha}^{~\gamma}
      +\omcirc_{\alpha\gamma}\,\Gamma_{a~\beta}^{~\gamma}
 \right)=0
\ee
This equation tells us that the matrix 
$\Gamma_a\equiv(\Gamma_{a~\beta}^{~\alpha})$
is an element of $\spl$. As a consequence, the coefficients
\bel{3.36}
\Gamma_{a\alpha\beta}\equiv-\omcirc_{\alpha\gamma} \Gamma_{a~\beta}^{~\gamma}
\ee
have the symmetry $\Gamma_{a\alpha\beta} =\Gamma_{a\beta\alpha}$.

The $\Mp$--covariant derivative of arbitrary tensorial objects
$\chi^{x\cdots \alpha\cdots}_{y\cdots \beta \cdots}$
is of the form $\nabla_a=\partial_a+\tfrac{i}{\hbar}\Gamma_a$
with 
\bel{3.37}
\Gamma_a=\tfrac{\hbar}{2} \,\Gamma_{a\alpha\beta}\,\Sigma^{\alpha\beta}
\ee
Here $\Sigma^{\alpha\beta}$ can be in any representation of
$\Sp$ or $\Mp$.
The $\Sigma$'s acting on the spinor indices are given by \rf{1.15};
the generators in the vector representation and its dual are 
such \cite{metino} that each upper frame index contributes a
term with a left--matrix action of $\Gamma_a=(\Gamma_{a~\beta}^{~\alpha})$,
and each lower index a term where $-\Gamma_a$ acts from the right.
Restricting ourselves to spinors, eq.\rf{3.37} is a special example
of the spin connection which we introduced in the previous subsection.
It assumes values in the Lie algebra $\mpl$ rather than the full 
algebra $\cg$ of all hermitian operators.

Under a local frame rotation a spinor transforms as 
\bel{3.38}
\delta_F \psi^x=-\tfrac{i}{2} \kappa_{\alpha\beta}(\phi)\,
\Sigma^{\alpha\beta x}_{~~~y}\, \psi^y
\ee
and arbitrary multispinors obey \rf{3.11} with
$U(\phi)=1-i\varepsilon(\phi)/\hbar$ where
\bel{3.39}
\varepsilon(\phi) \equiv \tfrac{\hbar}{2}\,
\kappa_{\alpha\beta}(\phi)\, \Sigma^{\alpha\beta}
\ee
This entails that the $\gamma$--matrices and the $\Mp$--generators
do not change at all if the transformations acting on the vector
indices are included: $\delta_F \gamma^{ax}_{~~y}=0$,
$\delta_F \Sigma^{\alpha\beta x}_{~~~y}=0$. From \rf{3.34}, \rf{3.36}
and the  transformation law for the vielbeins it follows that
\bel{3.40}
\delta_F \Gamma_{a\alpha\beta} 
=
\partial_a \kappa_{\alpha\beta} +\kappa_{\alpha\gamma}
\omcirc^{\gamma\delta}\Gamma_{a\beta\delta}
+\kappa_{\beta\gamma}\omcirc^{\gamma\delta}
\Gamma_{a\alpha\delta}
\ee
The change of the spin connection is 
$\delta_F\Gamma_a=\tfrac{\hbar}{2}(\delta_F \Gamma_{a \alpha \beta})
 \Sigma^{\alpha\beta}$.
It is easy to check that this  coincides precisely with 
the Yang--Mills gauge transformation \rf{3.8} if the generator
$\varepsilon$ is given by \rf{3.39}.

Under a coordinate transformation on $\cm_{2N}$, $\Gamma_{ab}^{c}$
transforms like a Christoffel symbol and $\Gamma_{a\alpha\beta}$
as a vector. Hence the operator \rf{3.37} satisfies \rf{3.8a}.

Up to this point the discussion has a certain similarity with
the theory of spinors on curved space--times. Now we come to a
special feature of symplectic geometry not shared by Riemannian
geometry. Typically, general coordinate transformations can be
used to transform a metric $g_{\mu\nu}$ to the flat metric
$\eta_{\mu\nu}$ in a single point only. In the symplectic case, 
the situation is much more favorable: according to 
Darboux's theorem there exist local coordinates with respect
to which $\omega_{ab}=\omcirc_{ab}=const$ can be achieved
in an extended domain. Using such Darboux local coordinates
$\phi^a$, the symplectic structure $\omega_{ab}$ is of the 
canonical form \rf{3.28} at all points of $\cm_{2N}$ covered
by the chart to which the $\phi^a$'s belong.

If $\omega_{ab}=\omcirc_{ab}$ then eq.\rf{3.30} tells us that
the vielbein coefficients must be chosen such that 
$[e^\alpha_a]\in \Sp$.
The simplest and most natural choice is
\bel{3.41}
e^\alpha_a =\delta^\alpha_a
\qquad , \qquad
e_\alpha^a =\delta_\alpha^a
\ee
In the following we shall always employ this specific
vielbein when we use Darboux coordinates. No distinction
between frame and coordinate indices is necessary then,
and we shall use the uniform notation $a,b,c, \cdots$ in either
case. From now on, and as in the previous sections, $\omega_{ab}$
stands for the constant canonical matrix \rf{2.3}.
Furthermore, eq.\rf{3.34} shows that for the vielbein \rf{3.41} 
the connection coefficients $\Gamma_{a~\beta}^{~\alpha}$
and $\Gamma^c_{ab}$ become identical. Thus, in Darboux
coordinates,
\bel{3.42}
\Gamma_{abc}=-\omega_{ad}\,\Gamma^d_{bc}
\ee
and
\bel{3.43}
\Gamma_a =\tfrac{\hbar}{2}\, \Gamma_{abc}\, \Sigma^{bc}
\ee
Using eq.\rf{3.32} together with the fact that the 
connection has vanishing torsion shows that $\Gamma_{abc}$
is symmetric in all three indices: 
$\Gamma_{abc}=\Gamma_{(abc)}$. In Darboux coordinates,
the $\Mp$--covariant derivative reads explicitly:
\bel{3.43a}
\bay
\nabla_a \chi^{x\cdots b \cdots}_{y\cdots c \cdots}
&=&\dis\partial_a \chi^{x\cdots b \cdots}_{y\cdots c \cdots}
+\tfrac{i}{2}\Gamma_{ade}
\left[
  \left(\Sigma^{de}\right)^x_{~z}\chi^{z\cdots b\cdots}_{y\cdots c \cdots}
       -\chi^{x\cdots b\cdots}_{z\cdots c\cdots}
  \left(\Sigma^{de}\right)^z_{~y}
\right]
\\[.3cm]
&&
\dis
+\Gamma^b_{ae} \,\chi^{x\cdots e\cdots}_{y\cdots c\cdots}
-\Gamma^{e}_{ac}\, \chi^{x\cdots b\cdots}_{y\cdots e\cdots}+\cdots
\eay
\ee    
In particular, $\nabla_a\gamma^b=0$ and $\nabla_a\Sigma^{bc}=0$.
The curvature of $\nabla_a$ is given by
\bel{3.43b}
\Omega_{ab}=\tfrac{\hbar}{2}\,R_{abcd}\, \Sigma^{cd}
\ee
The tensor $R_{abcd}$ is antisymmetric in $a$ and $b$ but
symmetric in $c$ and $d$. Thanks to the simple choice of
the vielbein, it is directly related to the Riemann tensor:
\bel{3.43c}
R_{ab~d}^{~~c} =-\omega^{ce} R_{abed}
\ee
In our conventions,
\bel{3.43d}
R_{ab~d}^{~~c} =\partial_a\Gamma^c_{bd}
-\partial_b \Gamma^c_{ad} +\Gamma^c_{ae}\Gamma^e_{bd}
-\Gamma^c_{be}\Gamma^e_{ad}
\ee

It is an important question which subgroup of the group
of diffeomorphisms and local frame rotations is left
unbroken after fixing the vielbein \rf{3.41}.
From the general transformation laws
\bel{3.44}
\bay
\delta_Ce^\alpha_a &=& \dis h^b\partial_be^\alpha_a
                      +\partial_ah^b e^\alpha_b \\
\delta_Fe^\alpha_a &=& \dis \omcirc^{\alpha\beta}
                       \kappa_{\beta\gamma}\,e^\gamma_a
\eay
\ee
it is clear that $e^\alpha_a=\delta^\alpha_a$ is not stable
under either of these transformations. However, if both the
vector field $h^a$ and the parameters $\kappa_{ab}$ are 
defined in terms of the same generating function $H(\phi)$ by
\bel{3.45}
h^a=\omega^{ab}\partial_b H 
\qquad , \qquad
\kappa_{ab}=-\partial_a\partial_b H
\ee
then the combined transformation
$\delta_{CF}\equiv \delta_C+\delta_F$
leaves the vielbein \rf{3.41} invariant:
\bel{3.46}
\delta_{CF} e^\alpha_a=0
\qquad {\rm for} \qquad
e^\alpha_a=\delta^\alpha_a
\ee
Therefore the stability group of the fixed vielbein,
under which the theory continues to be covariant,
is the group of symplectic diffeomorphisms of $\cm_{2N}$,
accompanied by appropriate frame rotations which are
completely determined by the generating function
of the diffeomorphism.

From \rf{3.8a} and \rf{3.38} we find the action of $\delta_{CF}$
on spinors:
\bel{3.45a}
\delta_{CF} |\psi \ra_\phi =h^a\partial_a |\psi\ra_\phi
+\tfrac{i}{2} \partial_a\partial_b H\, \Sigma^{ab} |\psi\ra_\phi
\ee
Remarkably enough, this equals precisely the Lie derivative
$\ell_H|\psi\ra_\phi$ of eq.\rf{2.12}. For arbitrary
multispinors, the continuous frame rotations which 
occur during the Lie--dragging governed by
$-\partial_t\chi =\ell_H \chi= \delta_{CF}\chi$ are
precisely those which are needed in order to preserve
the canonical form of the vielbein.

\subsection{ Parallel transport}
Let us fix a generic \cg--valued connection $\Gamma_a$
and an arbitrary smooth, oriented curve
on $\cm_{2N}$, $\cc=\cc(\phi_2, \phi_1)$, connecting
the points $\phi_2$ and $\phi_1$. We choose a parametrization
$\phi^a(s)$, $s\in[0,1]$, with
$\phi^a(0)=\phi^a_1$ and $\phi^a(1)=\phi^a_2$. Let us pick a
vector $|\psi\ra_{\phi_1}$ which lives in the fiber $\cv_{\phi_1}$
located at the initial point of the trajectory. We can parallel
transport this state to all Hilbert spaces $\cv_{\phi(s)}$
along the curve $\phi^a(s)$. The family 
$|\psi \ra_{\phi(s)}\in \cv_{\phi(s)}$ of parallel 
Hilbert space vectors satisfies 
$\dot\phi^a\nabla_a |\psi\ra_{\phi(s)}=0$ or
\bel{3.46a}
\frac{d}{ds}|\psi\ra_{\phi(s)}=
-\tfrac{i}{\hbar}\dot\phi^a(s) \,\Gamma_a\left(\phi(s)\right)
\,|\psi \ra_{\phi(s)}
\ee
where the dot means $d/ds$. (We avoid the notation $t$ for
the parameter here, because in general $\phi^a(s)$ is not 
related to a hamiltonian flow.) 
Eq.\rf{3.46a} can be solved formally in terms of a path--ordered
exponential,
\bel{3.47}
|\psi\ra_{\phi(s)} =
V\left[\cc\left(\phi(s), \phi_1\right)\right]
|\psi\ra_{\phi_1}
\ee
with ( $P$ denotes the path ordering operator)
\bel{3.48}
V\left[\cc\left(\phi(s), \phi_1\right)\right]
=P \exp\left[
-\tfrac{i}{\hbar}\int_0^s ds^\prime \,
\dot\phi^a(s^\prime) \Gamma_a\left(\phi(s^\prime)\right)\right]
\ee
In particular, $|\psi\ra_{\phi_2}=V[\cc(\phi_2,\phi_1)]|\psi\ra_{\phi_1}$.
The vector $|\psi\ra_{\phi_2}$ depends on $\cc$ in general.
The obvious generalization for multispinors is
\bel{3.49}
\chi_{\cal C}(\phi_2) ^{x\cdots}_{y\cdots}
=V\left[\cc(\phi_2, \phi_1)\right]^x_{~v}
\,V^\da\left[\cc(\phi_2, \phi_1)\right]^w_{~y}\,\cdots\, 
\chi(\phi_1)^{v\cdots}_{w\cdots}
\ee
We write $\chi_{\cal C}(\phi_2)$ for the transported multispinor
in order to indicate its path dependence. If $\Gamma_a$ is
a flat connection, $\Omega_{ab}(\Gamma)=0$, the
parallel transport is path--independent and every vector
in the ''reference Hilbert space'' at $\phi_1$ gives rise
to a spinor field over the entire phase--space:
$|\psi\ra_\phi=V[\cc(\phi, \phi_1)]|\psi\ra_{\phi_1}$.
It is ''covariantly constant": $\nabla_a|\psi\ra_\phi=0$.

Because the connection is a hermitian operator, the
parallel transport operator $V$ is unitary, and inner
products are preserved under parallel transport. One easily
verifies that if $\Gamma_a$ is gauge transformed according
to \rf{3.7}, $V$ changes as
\bel{3.50}
V\left[\cc(\phi_2, \phi_1)\right]^\prime
=U(\phi_2)\,V\left[\cc(\phi_2, \phi_1)\right]\,U(\phi_2)^\da
\ee
As a consequence, bilinears of the type
$
{}_{\phi_2}\la\chi| V\left[\cc(\phi_2, \phi_1)\right] |\psi\ra_{\phi_1}
$
are invariant under local frame rotations. 

\mysection{Split Symmetry and Exact Quantum
           Mechanics}
In the following we return to the question raised at the
end of section 2: given the fact that semiclassical
quantum mechanics has a natural interpretation in terms
of world--line spinors, what is the corresponding many--Hilbert
space description of full--fledged, exact quantum mechanics?
Rather than using the standard single--Hilbert space formulation
as a guide line we postulate an independent physical
principle, invariance under the background--quantum split
symmetry, and show a posteriori that this is precisely
equivalent to the ordinary quantization rules.
\subsection{The background--quantum split symmetry}
In classical mechanics, a solution of Hamilton's equation
infinitesimally close to a solution $\phi^a_\rcl(t)$ is given
$\phi^a_\rcl(t)+c^a(t)$ where $c^a(t)$ is a Jacobi field.
In semiclassical quantum mechanics, the $\hat\phi^a$--expectation 
value for a particle in the vicinity of $\phi^a_\rcl$ reads
$\phi^a_\rcl(t)+\bar\eta(t)\hat\varphi^a\eta(t)$.
Let us consider the situation at a fixed time $t$ when the
particle is at the point $\phi^a_\rcl(t)\equiv\phi^a$.
In classical mechanics we parametrize a point $\Phi^a$ which
is infinitesimally close to $\phi^a$ according
to $\Phi^a=\phi^a+\xi^a$ where $\xi\in T_\phi\cm_{2N}$ is 
a vector in the tangent space at $\phi$, and in semiclassical
quantum mechanics one parametrizes
$\Phi^a=\phi^a+{}_\phi\la\psi|\hat\varphi^a|\psi\ra_\phi$
in terms of a Hilbert space vector $|\psi\ra_\phi\in\cv_\phi$.

In a fully quantum mechanical situation a particle is not forced
to stay in the vicinity of any classical trajectory and 
therefore ${}_\phi\la \psi |\hat\varphi^a|\psi\ra_\phi$ is
not  small 
relative  to the classical contribution $\phi^a$. Thus, in
order to understand exact quantum mechanics in our framework, 
we need a tool to handle situations when $\Phi^a$ and $\phi^a$
are not close to each other.

In the classical case there exists a familiar method
of parametrizing points $\Phi^a$ in a (non--infinitesimal)
neighborhood of $\phi^a$ in terms of vectors $\xi\in T_\phi\cm$.
It is based on geodesics and is at the heart of the 
Riemann normal coordinate expansion \cite{frie,howe,sil}.
Let $\Gamma^c_{ab}$ be an arbitrary connection on $\cm_{2N}$, and
let $\Phi^a(s)$ denote the solution of the pertinent geodesic
equation with the initial point 
$\Phi^a(s=0)=\phi^a$ and the initial velocity $\dot\Phi^a(s=0)=\xi^a$.
We set $\Phi^a\equiv\Phi^a(1)$ for the point which is visited by
the trajectory at the ''time'' $s=1$. In this manner one can
establish a
diffeomorphism between a ball in
$T_\phi\cm_{2N}$, centered at $\xi=0$, and a neighborhood of $\phi$
in $\cm_{2N}$. Points $\Phi$ in this neighborhood are 
characterized by the initial velocity $\xi$
of the geodesic connecting them to the reference point $\phi$.
From the geodesic equation one finds explicitly :
\bel{4.1}
\begin{array}{c}
\dis
\Phi^a =\phi^a+\xi^a-\sum^\infty_{n=2} 
\frac{1}{n!}\, \Gamma^a_{b_1\cdots b_n}(\phi)
\,\xi^{b_1}\cdots \xi^{b_n}
\\[.3cm]
\dis
\Gamma^a_{b_1\cdots b_n}(\phi)\equiv
\tilde\nabla_{(b_1}\cdots\tilde\nabla_{b_{n-2}}
\Gamma^a_{b_{n-1} b_n)}(\phi)
\eay
\ee
The ''covariant derivative'' $\tilde\nabla^a$ is to be taken
with respect to the lower indices only. When applied to a flat
space with $\Gamma^c_{ab}=0$, eq.\rf{4.1} collapses to 
\bel{4.2}
\Phi^a=\phi^a+\xi^a
\ee
and the vectors $\xi$ in a single tangent space are sufficient
to parametrize the entire manifold. 
($\xi$ is not infinitesimal in \rf{4.2}!)

The parametrization \rf{4.1} is subject to a rather restrictive
consistency condition. In the simple example \rf{4.2} it is 
obvious that  for an arbitrary vector $\varepsilon^a$ the
pairs $(\phi^a,\xi^a)$ and $(\phi^a+\varepsilon^a, \xi^a-\varepsilon^a)$,
respectively, describe the same point $\Phi^a$. Stated differently,
$\Phi^a$ is invariant under the {\it split symmetry}
$\delta_S\phi^a=\varepsilon^a$, $\delta_S\xi^a=-\varepsilon^a$.
This is a rather trivial symmetry transformation for a flat 
space, but in the general case it involves a complicated
nonlinear transformation \cite{sil}:
\bel{4.3}
\bay
\delta_S\phi^a&=&\varepsilon^a\\
\delta_S\xi^a&=& F^a_b(\phi, \xi)\varepsilon^b
\eay
\ee
The functions $F^a_b$ must be determined from the requirement
that $\delta_S\Phi^a=0$.

We start the investigation of the quantum case with the
example of the flat phase--space $\cm_{2N}=\R^{2N}$. In 
this case
\bel{4.4}
\Phi^a=\phi^a +{}_\phi\la\psi|\hat\varphi^a|\psi\ra_\phi
\ee
provides a globally valid parametrization of phase--space
in terms of Hilbert space vectors. (The expectation
value in \rf{4.4} is not infinitesimal.) With the reference
point $\phi^a$ kept fixed, every $|\psi\ra_\phi\in\cv_\phi$
gives rise to a well--defined point $\Phi^a$. 
Conversely,
every $\Phi^a$ can be represented in the form \rf{4.4},
but there are infinitely many Hilbert space vectors which
achieve this. Let us ask how $|\psi\ra_\phi$ must
change in order to compensate for an infinitesimal change of
the reference point, $\phi^a\to\phi^a+\delta\phi^a$. 
We make the following ansatz for the ''background--quantum
split symmetry''
\bel{4.5}
\bay
\delta_S\phi^a&=&\varepsilon^a \\
\delta_S|\psi\ra_\phi&=&\dis
-\,\tfrac{i}{\hbar}\,\varepsilon^a\,
\tilde\Gamma_a(\phi) \,|\psi\ra_\phi
\eay
\ee
and determine the operators $\tilde\Gamma^a$ in such a way
that $\delta_S\Phi^a=0$. Applying \rf{4.5} to \rf{4.4}
yields the condition
\bel{4.6}
\left[\tilde\Gamma_a,\hat\varphi^b\right]=i\hbar\,\delta_a^b
\ee
if one assumes that $|\psi\ra_\phi$ is normalized to unity. 
Its general solution reads
\bel{4.7}
\tilde\Gamma_a(\phi)=\omega_{ab}\,\hat\varphi^b+\alpha_a(\phi)
\ee
where $\alpha_a$ is an arbitrary $c$--number valued one--form.
Thus we managed to associate to $\phi^a$,
regarded as a classical observable, a family of operators
(labeled by the points of phase--space)
\bel{4.8}
\co_{\phi^a}(\phi)=\phi^a+\hat\varphi^a
\ee
such that 
${}_\phi\la\psi| \co_{\phi^a}(\phi)|\psi\ra_\phi$
is independent of $\phi$ provided the states $|\psi\ra_\phi$
at different points $\phi$ are related by \rf{4.5}.

Returning now to an arbitrary curved phase--space, we expect
that the RHS of eq.\rf{4.8} has to be modified by terms involving
products of $\hat\varphi$'s. Clearly
eq.\rf{4.8} is a kind of quantum analogue of \rf{4.2}, and the
r\^ole of $\xi^a$ is played by the operators $\hat\varphi^a$ now.
Rather than looking at $\phi^a$ only, we shall consider a 
general classical observable, i.e., a real--valued function
$f=f(\phi)$, and associate a family of operators
\bel{4.9}
\co_f(\phi) =f(\phi) +\Delta \co_f(\phi)
\ee
to it such that 
${}_\phi\la\psi| \co_f(\phi)|\psi\ra_\phi$
is independent of $\phi$. In \rf{4.9}, $\Delta \co_f(\phi)$
stands for a power--series in $\hat\varphi^a$ which
starts with a linear term. 

Neither the form of $\co_f(\phi)$ nor the prescription for
transporting the $|\psi\ra$'s is known a priori. We shall
keep the ansatz \rf{4.5} for the split symmetry also in 
the general case, but \rf{4.7} will have to be generalized.
Assume $\tilde\Gamma_a$ is given, then we can transport any
vector $|\psi\ra_\phi$ from $\phi$ to $\phi+\delta\phi$:
\bel{4.7a}
\delta_S|\psi\ra_\phi \equiv
|\psi\ra_{\phi+\delta\phi}
-|\psi\ra_{\phi}=\varepsilon^a\,\partial_a |\psi\ra_{\phi}
+O(\varepsilon^2)
\ee
This means that the function
$\phi\mapsto|\psi\ra_{\phi}$ is constrained by the condition
\bel{4.8a}
i\hbar\,\partial_a|\psi\ra_{\phi}=
\tilde\Gamma_a(\phi)\,|\psi\ra_{\phi}
\ee
If we perform local frame rotations of the form \rf{3.5}
both in $\cv_\phi$ and in $\cv_{\phi+\delta\phi}$, 
we observe that the operators $\tilde\Gamma_a$ transform
precisely according to \rf{3.7}.
This means that $\tilde\Gamma_a$ is a spin connection
of the type introduced in section 3.1. Eq.\rf{4.8a} tells
us that the spinor field $\psi^x(\phi)=\la x|\psi\ra_\phi$
should be ''covariantly constant''
with respect to $\tilde\Gamma_a$: $D_a|\psi\ra_\phi=0$.
We shall use the notation
\bel{4.9a}
D_a\equiv\partial_a+\tfrac{i}{\hbar}\,\tilde\Gamma_a
\equiv \nabla_a(\tilde\Gamma)
\ee
for the covariant derivative formed from $\tilde\Gamma_a$.
The split transformation $\delta_S$ is nothing but an
infinitesimal parallel transport from $\phi$ to $\phi+\delta\phi$.
If we require invariance under the split symmetry \rf{4.5}, i.e.,
that 
${}_\phi\la\psi| \co_f(\phi)|\psi\ra_\phi$
is independent of $\phi$, it follows that $\co_f$, too, must
be covariantly constant (or a ''flat section''):
\bel{4.10}
\partial_a\co_f +\tfrac{i}{\hbar}
\left[\tilde\Gamma_a, \co_f\right]
\equiv D_a \co_f=0
\ee
A necessary condition for the integrability of this equation
is that 
$D_{[a}D_{b]}\co_f=0$, or
\bel{4.11}
\left[\Omega_{ab}(\tilde\Gamma), \co_f(\phi)\right]=0
\ee
We conclude that $\tilde\Gamma$ must be an abelian connection.
By definition \cite{fedbook}, an abelian connection has a 
curvature which commutes with all other operators. This
happens if $\Omega_{ab}(\tilde\Gamma)$ equals a $c$--number valued
2--form times the unit operator on \cv: 
$\Omega_{ab} (\tilde\Gamma)=\beta_{ab}(\phi)$.
As a consequence, the resulting parallel transport operator
$V[\cc]$ for a closed curve $\cc$ is simply a phase factor
times the unit operator, i.e., the holonomy group of $\tilde\Gamma$
is $U(1)$ rather than the full group $G$. For $\cc$ an infinitesimal
parallelogram, for instance, one has 
$V[\cc]=1-\frac{i}{\hbar}\beta_{ab}d\phi^ad\phi^b$.

The analogous integrability condition of the equation for 
states, $D_a|\psi\ra_\phi=0$, is more restrictive. It requires
the connection to be flat: $\Omega_{ab}(\tilde\Gamma)=0$.
If this condition is satisfied, we can parallel transport a state
$|\psi\ra_\phi$ around a closed loop $\cc$ and, at least locally,
we are guaranteed to get back precisely the original state after
having completed the circuit. If, instead, the connection is not
flat but only
abelian, $|\psi\ra_\phi$ will in general
pick up a phase factor during the excursion around \cc.
For our purposes this is equally acceptable, because the phase
cancels in 
${}_\phi\la\psi| \co_f(\phi)|\psi\ra_\phi$.
Thus, in order to make sure that this expectation value
is independent of $\phi$, we shall only require
that $\tilde\Gamma_a$ is an abelian connection, and relax the
consistency condition at the level of $|\psi\ra_\phi$ to a
''consistency up to a phase''. In the next section we shall
see that an abelian connection can be found for any 
phase--space $\cm_{2N}$.

For  flat phase--space, $\tilde\Gamma_a$ is given in
eq.\rf{4.7}. Its curvature 2--form reads
\bel{4.12}
\Omega(\tilde\Gamma)=\omega+d\alpha
\ee
As it should be, this connection is abelian: both the
symplectic 2-form and $d\alpha$ are proportional to the
unit operator in \cv. Flat space is special in that 
$\omega$ is globally exact, i.e., $\omega=d\theta$ where
the symplectic potential $\theta=\theta_a d\phi^a$ has the
components
\bel{4.13}
\theta_a(\phi)=-\,\tfrac{1}{2}\,\omega_{ab}\phi^b
\ee
Hence setting $\alpha=-\theta$ gives us a flat connection:
\bel{4.14}
\tilde\Gamma_a(\phi)=\omega_{ab}
\left(\hat\varphi^b+\tfrac{1}{2}\phi^b\right)
\ee
This situation is not typical though. In the next section
we shall employ an algorithm which, in a ''canonical''
way, provides us with an abelian connection with 
curvature $\Omega(\tilde\Gamma)=\omega$. We could try
to modify $\tilde\Gamma$ by adding a $c$--number valued
1--form $\alpha$;
this leads to the curvature $\Omega(\tilde\Gamma+\alpha)=\omega+d\alpha$.
But generically $\omega$ can not be written as $d\theta$ in terms of a
globally well defined 1--form $\theta$. Therefore it is not 
possible then to obtain a flat connection by setting $\alpha=-\theta$.

To summarize: To every classical observable $f$ and to every
point $\phi$ of phase--space we have associated an operator
$\co_f(\phi):\,\cv_\phi\to\cv_\phi$. In each fiber
$\cv_\phi$ we can compute expectation values of the form
\bel{4.15}
{}_\phi\la\psi| \co_f(\phi)|\psi\ra_\phi
=f(\phi)+
{}_\phi\la\psi| \Delta\co_f(\phi)|\psi\ra_\phi
\ee
We have postulated that this expectation value
should be the same in all fibers, i.e., independent of $\phi$.
This requires $\co_f(\phi)$, regarded as a function of $\phi$,
to be covariantly constant with respect to an arbitrary abelian
connection $\tilde\Gamma$ and $|\psi\ra_\phi$ to be covariantly
constant up to a phase. Changing the point $\phi$ in \rf{4.15} 
means changing the individual contributions of the classical
''background'' term $f(\phi)$ and of the quantum mechanical
$\Delta\co_f$--expectation value. The position
independence of their sum is what we mean  by 
''background--quantum split symmetry''.

\subsection{Abelian connections}
The implementation of the split symmetry involves two steps:
first, find an abelian connection $\tilde\Gamma$, and second,
construct multispinor fields (states, operators, ...) which
are covariantly constant with respect to $\tilde\Gamma$, possibly
modulo a phase.
We shall employ a method which was proposed by Fedosov
\cite{fed,fedbook} and which allows for an iterative 
construction of $\tilde\Gamma$ on any symplectic manifold.

We start by fixing an arbitrary symplectic connection.
We write $\Gamma$ for this connection and $\nabla_a\equiv\nabla_a(\Gamma)$
for its covariant derivative, and continue to use the notation
$\tilde\Gamma$ and $D_a\equiv\nabla_a(\tilde\Gamma)$ for the abelian
connection and its covariant derivative. We make the ansatz
\bel{4.16}
\tilde\Gamma=\Gamma+\lambda
\ee
where $\lambda=\lambda_a(\phi)\, d\phi^a$ is some globally defined
1--form. While $\Gamma\in \mpl$, $\lambda$ may assume
values in the larger algebra \cg. The curvature forms of
$\Gamma$ and $\tilde\Gamma$ are related by
\bel{4.17}
\Omega(\tilde\Gamma)=\Omega(\Gamma) +\nabla\lambda
+\tfrac{i}{\hbar}\, \lambda^2
\ee
It will turn out convenient to separate off from $\lambda_a$ the
flat space solution \rf{4.7}:
\bel{4.18}
\lambda \equiv \,\omega_{ab}\,\hat\varphi^b\,d\phi^a+r
\ee
Inserting \rf{4.18} into \rf{4.17} one finds after some
calculation which makes essential use of the fact that 
$\Gamma$ is symplectic:
\bel{4.19}
\Omega(\tilde\Gamma)=\omega+\Omega(\Gamma)
-\delta r +\nabla r +\tfrac{i}{\hbar} \,r^2
\ee
Here we introduced the linear map $\delta$. It acts on 
operator--valued differential forms $F$ according to 
\bel{4.20}
\delta F =d\phi^a \,\hat\partial_a F
\ee
with 
\bel{4.21}
\hat\partial_a F= -\tfrac{i}{\hbar}\, \omega_{ab} 
\left[\hat\varphi^b, F\right]
\ee

At this point some technical remarks are in order. 
The connection $\tilde\Gamma$ assumes values in the
Lie algebra $\cg$
consisting of all hermitian operators on \cv. More 
precisely, we define $\cg$ to be generated by completely
symmetrized products of $\hat\varphi$'s:
\bel{4.22}
\hat\varphi^{(a_1}\cdots\hat\varphi^{a_n)}
\qquad , \qquad n=0,1,2,3,\cdots
\ee
These operators are Weyl--ordered. Their commutator
algebra is a (generalization of the) $W_\infty$--algebra
\cite{gelf,winf}.
The quadratic generators 
$\hat\varphi^{(a} \hat\varphi^{b)}=\hbar\,\Sigma^{ab}$
form the subalgebra of \mpl.

When applied to a symmetrized monomial \rf{4.22},
$\hat\partial_a$ acts formally like a differentiation
$\partial/\partial\hat\varphi^a$, i.e.,
if $g(\varphi)$ is an analytic $c$--number function
and we associate the operator
\bel{4.23}
g(\hat\varphi)\equiv
\sum^\infty_{n=0} \frac{1}{n!}
\,\partial_{a_1}\cdots\partial_{a_n} g(0)\,
\hat\varphi^{(a_1}\cdots \hat\varphi^{a_n)}
\ee
to it, then 
$\hat\partial_a g(\hat\varphi)=(\partial g/ \partial\varphi^a)(\hat\varphi)$.
The operation $\hat\partial_a$ has to be carefully distinguished 
from the partial derivative 
$\partial_a\equiv \partial/\partial\phi^a$.

Frequently we shall deal with monomials of the form
\bel{4.24}
\hat{F}_{mn} \equiv \hbar^m
\hat\varphi^{(a_1}\cdots \,\hat\varphi^{a_n)}
\ee
An important concept is their degree which is defined
as ${\rm deg}(\hat F_{mn})=2m+n$, i.e., 
${\rm deg}(\hbar)=2$ and ${\rm deg}(\hat\varphi^a)=1$. 
For a product of two monomials \rf{4.24}, the degree is
additive:
\bel{4.25}
{\rm deg}
(F_1 \,F_2)={\rm deg}(F_1)+{\rm deg}(F_2)
\ee
A priori, $F_1 F_2$ is not totally symmetrized if $F_1$ and $F_2$
are, but it can be expanded in the basis \rf{4.22} by repeatedly
applying the canonical commutation relations \rf{1.9}. Each time
\rf{1.9} is used, two $\hat\varphi$'s are removed and 
one factor of $\hbar$ is added, thus leaving the degree unchanged.
Hence all terms generated in the process of Weyl ordering
$F_1F_2$ have the same degree, ${\rm deg}(F_1) +{\rm deg}(F_2)$.

In the spirit of a semiclassical expansion, the degree is the
{\it total} power of $\hbar^{1/2}$ contained in a given monomial.
In fact, the $\hat\varphi$'s satisfy the commutation relation
\rf{1.9} which involves an explicit factor of $\hbar$;
hence $\hat\varphi^a$ itself is of order $\hbar^{1/2}$.
The $\gamma$--matrices satisfy an $\hbar$--independent
Clifford algebra instead, and are of degree zero therefore.
Reexpressing $\hat F_{mn}$ in terms of $\gamma$--matrices, the
degree coincides with the explicit power of $\hbar^{1/2}$:
\bel{4.26}
\hat F_{mn}=
2^{-n/2} \,\hbar^{(2m+n)/2}
\gamma^{(a_1}\cdots \gamma^{a_n)}
\ee

Let us return to eq.\rf{4.19}. It is clear that if we manage
to find an operator valued 1--form $r\equiv r_a\,d\phi^a$ in 
such a way that
\bel{4.27}
\delta r=\Omega(\Gamma) +\nabla r +\tfrac{i}{\hbar}\, r^2
\ee
then
\bel{4.28}
\tilde\Gamma_a=\Gamma_a+\omega_{ab}\,\hat\varphi^b + r_a
\ee
is an abelian connection with 
\bel{4.29}
\Omega_{ab}\left(\tilde\Gamma \right)=\omega_{ab}
\ee
It is here that the work of Fedosov \cite{fed, fedbook} 
comes in. He showed that \rf{4.27} can be solved on any
symplectic
manifold and he developed an iterative procedure for 
calculating $r$. If one wants to compute $r$ only up to
terms of some fixed degree only a finite number of iterations
is needed. Some details of this method are given in the 
Appendix. Here we only quote the first two terms of the result:
\bel{4.30}
r(\phi,\hat\varphi) =
-\tfrac{1}{8}\,R_{abcd}\, 
\hat\varphi^{(b} \hat\varphi^c
\hat\varphi^{d)} \,d\phi^a
+
\tfrac{1}{40}\,\nabla_b R_{aecd}\,
\hat\varphi^{(a} 
\hat\varphi^b
\hat\varphi^c
\hat\varphi^{d)}\, d\phi^e\, +\,O(5)
\ee
$R\equiv \Omega(\Gamma)$ is the curvature of the
symplectic connection, and ''$O(5)$'' stands for terms
of degree 5 and higher.

Our second task is the construction of covariantly
constant fields $\chi$, in particular of the operators
$\co_f(\phi)$.
Given $\tilde\Gamma$, Fedosov's method can also be used
to solve the equation $D_a\co_f=0$ by iteration.
For any function $f$, this equation has a unique solution if
one requires that the term in $\co_f(\phi)$ 
proportional to the unit operator is
precisely $f(\phi)$. In the appendix we show that
\bel{4.31}
\bay
\co_f(\phi) &=& \dis
f(\phi) +\hat\varphi^a\,\nabla_a f
+\tfrac{1}{2}\,\hat\varphi^{(a} \hat\varphi^{b)}\,\nabla_a\nabla_b f
+\tfrac{1}{6}\,\hat\varphi^{(a} \hat\varphi^{b}\hat\varphi^{c)}\,
 \nabla_a\nabla_b\nabla_c f
\\[.25cm]
&&\dis
+\,\tfrac{1}{24}\,R_{abcd}\, \omega^{be}\,\partial_e f\, 
\hat\varphi^{(a} \hat\varphi^{c}\hat\varphi^{d)}\,+\,O(4)
\eay
\ee
As an example, let us consider the observable $f(\phi)=\phi^a$
for $a$ fixed. With the notation $\tilde\nabla_a$ for the
''covariant derivative'' which acts on lower indices only, 
one finds
\bel{4.32}
\bay
\co_{\phi^a}(\phi) &=&\dis
\phi^a+\hat\varphi^a
-\tfrac{1}{2}\,\Gamma^a_{bc}\,
\hat\varphi^{(b} \hat\varphi^{c)}
-\tfrac{1}{3}\,
\tilde\nabla_{(b}\Gamma^a_{cd)}\,
\hat\varphi^{(b} \hat\varphi^c\hat\varphi^{d)}
\\[.3cm]
&&\dis
+\,\tfrac{1}{24} \,\omega^{ab}\, R_{bcde}\, 
\hat\varphi^{(c} \hat\varphi^d \hat\varphi^{e)}\,+\,O(4)
\eay
\ee

This expansion assumes a particularly simple form
if one uses normal Darboux coordinates. Every symplectic 
connection defines an essentially unique system of
Darboux local coordinates, the normal Darboux coordinates
\cite{fedbook}, with the property that at a given
point $\phi_0$ both $\Gamma_{abc}=0$ and
$\partial_{(e_1}\partial_{e_2}\cdots \partial_{e_n}
\Gamma_{abc)}=0$ for all $n=1,2,\cdots$.
From these equations one can deduce that 
$\partial_a\Gamma_{bcd}(\phi_0)=\frac{3}{4}R_{a(bcd)}(\phi_0)$,
and similar relations for higher partial derivatives of
$\Gamma_{abc}$. In terms of normal Darboux coordinates
centered at $\phi_0=\phi$, eq.\rf{4.32} becomes
\bel{4.33}
\bay
\co_{\phi^a}(\phi) &=& \dis
\phi^a+\hat\varphi^a-\tfrac{1}{24}\,\omega^{ab}\,R_{bcde}
\,\hat\varphi^{(c} \hat\varphi^d \hat\varphi^{e)} + O(4)
\\[.3cm]
&=& \dis
\phi^a+\hat\varphi^a-\tfrac{1}{6}\,
\tilde\nabla^a_{(b}\Gamma_{cd)}
\,
\hat\varphi^{(b} \hat\varphi^c \hat\varphi^{d)}+O(4)
\eay
\ee
This is precisely what one obtains from the classical expansion
\rf{4.1} if one replaces $\Phi^a\to \co_{\phi^a}$
and $\xi^a\to \hat\varphi^a\,$!
This correspondence will not persist at the higher orders,
though, since eventually the iteration will produce terms
with explicit powers of $\hbar$.\footnote{
For an interpretation of Fedosov's work in terms of 
a ''quantum deformed exponential map'' see ref.\cite{emm}.}

\subsection{Recovering exact quantum mechanics}
The split symmetry requires all fields $\co_f(\phi)$ to
be covariantly constant with respect to an abelian connection.
As the curvature $\Omega(\tilde\Gamma)=\omega$ is proportional
to the unit operator, the holonomy group of $\tilde\Gamma$, 
generated by the operators
$
V[\cc] = P \exp\bigg(-\tfrac{i}{\hbar} 
\int\limits_{\cal C}\tilde\Gamma\bigg)
$
for closed curves \cc, is reduced to a $U(1)$ subgroup of $G$.
The parallel transport around an infinitesimal parallelogram
yields $V[\cc]=1-\tfrac{i}{\hbar}\omega_{ab}d\phi^a d\phi^b$;
for arbitrary loops this integrates to 
$
V[\cc] = \exp\bigg(-\tfrac{i}{\hbar} 
\int\limits_{\cal D}\omega\bigg)
$
where $\cd$ is an arbitrary surface in $\cm_{2N}$ bounded by \cc,
$\partial\cd=\cc$. To begin with, let us focus on a single
chart of an atlas covering $\cm_{2N}$. Since $\omega$ is closed,
{\it locally} there exists an 1--form $\theta$ such that $\omega=d\theta$
and
$V[\cc] = \exp\bigg(-\tfrac{i}{\hbar} \int\limits_{\cal C}\theta\bigg)
$.
Let us suppose $\cc_1$ and $\cc_2$ are two open curves with 
the same endpoints. Then $\cc_1-\cc_2$ is closed and
\bel{4.36a}
V[\cc_1]\,V[\cc_2]^\da = \exp\bigg(-\tfrac{i}{\hbar}\!\! 
\int\limits_{{\cal C}_1-{\cal C}_2}\!\!\!\theta\bigg)
\ee
is a pure phase factor. Therefore, if one defines
\bel{4.37}
V[\cc(\phi,\phi_0)] \equiv \exp\bigg(-\tfrac{i}{\hbar}\!\! 
\int\limits_{{\cal C}(\phi ,\phi_0)}\!\!\!\theta\bigg)
\,\tau(\phi, \phi_0)
\ee
then the unitary operator $\tau$ depends only on the two points
but not on \cc. We can parallel transport an arbitrary
$(r,s)$--spinor from $\phi_0$ to $\phi$, and the only path
dependence
will reside in an overall phase factor:
\bel{4.38}
\bay
\chi_{\cal C}(\phi)^{x_1\cdots x_r}_{y_1\cdots y_s}
&=&\dis
\exp\bigg[ -\tfrac{i}{\hbar}(r-s)\!\!
\int\limits_{{\cal C}(\phi ,\phi_0)}\!\!\!\theta \bigg]
\\
&&\dis\qquad\times
\tau(\phi, \phi_0)^{x_1}_{~v_1}\cdots
\,\chi(\phi_0)^{v_1\cdots v_r}_{w_1\cdots w_s}
\,\tau^\da(\phi ,\phi_0)^{w_1}_{~y_1}\, \cdots
\eay
\ee
For $r=s$ the phase factor cancels, and in particular the
parallel transport of operators $(r=s=1)$ is path independent.
The most important example are the ''local observables''
$\co_f$:
\bel{4.39}
\co_f(\phi)=\tau(\phi,\phi_0)\,\co_f(\phi_0)\, \tau(\phi, \phi_0)^{-1}
\ee
For states, 
\bel{4.40}
|\psi_{\cal C}\ra_\phi =\exp \bigg( -\tfrac{i}{\hbar}\!\!\!\!
\int\limits_{{\cal C}(\phi ,\phi_0)}\!\!\!\!\theta \bigg)
\,\tau(\phi, \phi_0) \,|\psi\ra_{\phi_0}
\ee

Given the operator $\co_f(\phi_0)$ at a fixed point $\phi_0$,
eq.\rf{4.39} defines a $(1,1)$--spinor field on the entire
phase--space. Because of the path--dependent exponential, 
this is not the case for the vectors \rf{4.40}.
However, $|\psi\ra_{\phi_0}$ gives rise to well defined
spinor fields over a special class of submanifolds in $\cm_{2N}$.
In fact, every $2N$--dimensional symplectic manifold has 
certain $N$--dimensional submanifolds $\ck$, the Lagrangian
manifolds \cite{wood, lj}, with the property that for every
path $\cc$ in $\ck$ the integral $\int_{\cal C}\theta$ is invariant under
smooth deformations of $\cc$ which stay within $\ck$. Lagrangian
manifolds can be thought of as the image of 
configuration space under some canonical transformation. 
If $\phi_0\in \ck$, the parallel transport is locally integrable,
i.e., \rf{4.40} defines a single--valued spinor field $|\psi\ra_\phi$
on $\ck$ in a neighborhood of $\phi_0$. Globally, complications
might arise if two paths connecting $\phi$ to $\phi_0$ form a closed
curve in $\ck$ which cannot be smoothly contracted to a point while
remaining on \ck. This leads to topological quantization 
conditions \cite{wood} which we shall not discuss here.

As we pointed out at the beginning of this section, our aim 
is to show that semiclassical quantum mechanics plus invariance under
the background--quantum split symmetry implies
exact quantum mechanics. Eqs.\rf{4.39} and
\rf{4.40} are the main ingredients for establishing this 
equivalence. These two equations show that the physical
contents which is encoded in the local physical states
and operators related to the infinitely many Hilbert spaces
$\cv_\phi$ can be described by the states and operators
belonging to one single ''reference'' Hilbert space.
We choose this reference Hilbert space to be the fiber
$\cv_\phi$ at $\phi=\phi_0$ with $\phi_0$ an arbitrary
but fixed point of $\cm_{2N}$. Since the operators and 
states at any other point are related to those at $\phi_0$
by \rf{4.39} and \rf{4.40}, it does not matter at which
point transition matrix elements or expectation values
of physical observables are computed:
\bel{4.41}
\bay
{}_\phi\la \chi_{\cal C}| 
\psi_{\cal C}\ra_\phi
&=&
{}_{\phi_0}\la \chi| \psi\ra_{\phi_0}
\\
{}_\phi\la \chi_{\cal C}| \,
\co_f(\phi)\,|
\psi_{\cal C}\ra_\phi
&=&
{}_{\phi_0}\la \chi|\, 
\co_f(\phi_0)\,|
\psi\ra_{\phi_0}
\eay
\ee
Note that the LHS of these equations is actually independent
of $\cc$ because the path--dependent phase factor cancels.

In the standard formulation of quantum mechanics, the states
of a physical system are described by the vectors of a single
Hilbert space $\cv_{\rm QM}$ and observables 
by hermitian operators on $\cv_{\rm QM}$. The discussion above
suggests identifying $\cv_{\rm QM}$ with the reference
Hilbert space at $\phi_0$,
\bel{4.42}
\cv_{\rm QM}\, \cong \,\cv_{\phi_0}
\ee
and the quantum mechanical state vectors and observables
with the vectors $|\psi\ra_{\phi_0}\in \cv_{\phi_0}$
and operators $\co_f(\phi_0):\, \cv_{\phi_0}\to\cv_{\phi_0}$,
respectively. Let us make this identification more precise.

At the semiclassical level, the time evolution of multispinor
fields $\chi(\phi;t)$ is governed by the Lie--derivative
$\ell_H$. It remains to be understood how this type of dynamics
(''Lie--transport along the Hamiltonian flow'') is related to the
conventional Schr\"odinger or Heisenberg equation on $\cv_{\rm QM}$.
The unifying dynamical principle which, on the one hand,
reduces to the $\ell_H$--dynamics in the semiclassical
limit and on the other hand implies the ordinary 
Schr\"odinger equation when all Hilbert spaces are identified
is a time--dependent local frame
rotation with the generator $\co_H(\phi)$.
Let $\chi(\phi;t)$ be an arbitrary, not necessarily covariant
constant multispinor satisfying the equation of motion
\bel{4.43}
-\partial_t \chi(\phi;t) =\cl_H(\phi) \,\chi(\phi;t)
\ee
with
\bel{4.44}
\bay
\cl_H(\phi)\, 
\chi(\phi)^{x_1\cdots x_r}_{y_1\cdots y_s}
&\equiv&\dis 
\tfrac{i}{\hbar}\sum_{m=1}^r \co_H(\phi)^{x_m}_{~~z}\,
\chi(\phi)^{x_1\cdots z \cdots x_r}_{y_1\cdots y_s}
\\[.3cm]
&-&\dis
\tfrac{i}{\hbar}\sum_{m=1}^s 
\chi(\phi)^{x_1\cdots x_r}_{y_1\cdots z \cdots y_s}\,
\co_H(\phi)_{~y_m}^{z}
\eay
\ee
This equation of motion is the most general dynamical law,
based upon a local frame rotation, which is compatible with
the background--quantum split symmetry. We have to make sure
that if $D_a\chi=0$ is satisfied initially it still holds true
at any later time. This is the case if the generator
is annihilated by $D_a$, i.e., if it is of the form
$\co_H(\phi)$ for some function $H$.

Let us determine the semiclassical limit of the operator
\rf{4.44}, $\cl_H^{\rm scl}$. Due to the explicit factors of 
$\hbar$ in \rf{4.44} this amounts to disregarding all terms
of positive degree: $\cl_H =\cl_H^{\rm scl}+O(1)$. 
As a consequence, terms of degree 3 and higher may be omitted 
from $\co_H(\phi)$ and only the first three terms on the 
RHS of \rf{4.31} are relevant. Writing out the covariant
derivatives and using \rf{1.5} one finds
(with $h^a\equiv \omega^{ab}\partial_bH$):
\bel{4.45}
\tfrac{i}{\hbar}\co_H(\phi)
=
\tfrac{i}{\hbar}H(\phi)+\tfrac{i}{2}\partial_a\partial_bH(\phi)\Sigma^{ab}
-\tfrac{i}{\hbar}h^a\left[
\omega_{ab}\hat\varphi^b+\tfrac{\hbar}{2}\Gamma_{abc}\Sigma^{bc}
\right] +O(1)
\ee
It is interesting that
the terms in the square brackets are exactly
the degree--2 approximation of $\tilde\Gamma_a$.
In fact, because $r_a$ starts with a term of order 3, 
eq.\rf{4.28} with \rf{3.43} reduces to
\bel{4.46}
\tilde\Gamma_a =\omega_{ab}\,\hat\varphi^b
+\tfrac{\hbar}{2}\,\Gamma_{abc}\,\Sigma^{bc}+O(3)
\ee
We interpret the connection--term in \rf{4.45} as 
$\tfrac{i}{\hbar}\tilde\Gamma_a=D_a-\partial_a$. Thus,
\bel{4.47}
\bay
\cl_H^{\rm scl}
\,\chi(\phi)^{x_1\cdots x_r}_{y_1\cdots y_s}
&=&\dis
\left[h^a\partial_a-h^a D_a+\tfrac{i}{\hbar}(r-s)H\right]
\chi^{x_1\cdots x_r}_{y_1\cdots y_s}
\\[.3cm]
&&+\,\tfrac{i}{2}\partial_a\partial_b H
\left[\left(\Sigma^{ab}\right)^{x_1}_{~~z}\, \chi^{z\cdots}_{y_1\cdots}
-\chi^{x_1\cdots}_{z\cdots}\left(\Sigma^{ab}\right)^z_{~y_1}
\,\pm \cdots \right]
\eay
\ee
Comparison with \rf{2.13} shows that $\cl_H^{\rm scl}$ is closely
related to the Lie--derivative $\ell_H$:
\bel{4.48}
\cl_H^{\rm scl} \chi 
=\ell_H\chi -h^a D_a\chi +\tfrac{i}{\hbar}(r-s)H\chi
\ee
Obviously, $\cl_H^{\rm scl}$ and $\ell_H$ differ by terms involving
the classical Hamiltonian and $D_a\chi$. For multispinors 
with $r=s$, e.g., for operators $(r=s=1)$, the $H$--term
is absent. But also for other multispinors it can be removed
from the equation of motion \rf{4.43}, \rf{4.48}
in a trivial way: the field $\chi^\prime$ defined by
\bel{4.49}
\chi(\phi;t)
\equiv \exp\left[ -\tfrac{i}{\hbar}(r-s)H(\phi)t\right]
\chi^\prime(\phi; t)
\ee
satisfies the same equation without the $H$--term.
For $(1,0)$--spinors, eq.\rf{4.49} contains a 
generalization of the familiar dynamical phase $\exp(-iEt/\hbar)$.
The equivalence of $\cl_H^{\rm scl}$ and $\ell_H$ becomes complete
when they are applied to covariantly constant fields. Then
$D_a\chi=0$ and $\cl_H\chi=\ell_H\chi+O(1)$.

It is quite remarkable how this equality comes about. In contradistinction
to $\ell_H$, the operator $\cl_H$ effects a pure frame
rotation and no diffeomorphism. In fact, the partial derivatives
cancel between the first and the second term of \rf{4.48}.
It is only because the condition $D_a\chi=0$
couples displacements in $\cm_{2N}$ to rotations in $\cv_\phi$
that $\cl_H$ becomes effectively equivalent to a Lie derivative.

Under an infinitesimal frame rotation with parameters $\varepsilon$,
the connection changes by an amount $\delta_F\tilde\Gamma_a=D_a\varepsilon$.
In the case at hand, $\varepsilon\equiv \co_H(\phi)t$ is covariantly
constant, and the connection is invariant therefore:
$\delta_F\tilde\Gamma_a=0$.

The most important property of the equation of motion
\rf{4.43}, \rf{4.44} is its covariance with respect to
the split symmetry. It can be solved in terms of the local
evolution operators
\bel{4.50}
U(\phi;t)=
\exp\left[-\tfrac{i}{\hbar}\co_H(\phi)\,t\right]
\ee
They have the property
\bel{4.51}
U(\phi;t)\, \tau(\phi, \phi_0)
=\tau(\phi,\phi_0)\, U(\phi_0, t)
\ee
Applying this equation to a vector $|\psi(t=0)\ra_{\phi_0}$,
say, we see that it does not matter whether we first 
parallel--transport this vector from $\phi_0$ to $\phi$ and
time--evolve it there using $U(\phi;t)$ or if we first
time--evolve at $\phi_0$, using $U(\phi_0 ;t)$, and
parallel--transport afterwards:
\bel{4.52}
|\psi(t)\ra_\phi =U(\phi;t)\, |\psi(0)\ra_\phi
=V[\cc(\phi, \phi_0)]\, |\psi(t)\ra_{\phi_0}
\ee

The vectors $|\psi(t)\ra_\phi$ satisfy the local Schr\"odinger
equation
\bel{4.53}
i\hbar\,\partial_t |\psi(t)\ra_\phi=\co_H(\phi)\, |\psi(t)\ra_\phi
\ee
For different points $\phi$, these equations are related by
\rf{4.39}, \rf{4.40} and are equivalent therefore.
In particular, they are all equivalent to the
Schr\"odinger equation in the reference Hilbert space
$\cv_{\phi_0}\cong \cv_{\rm QM}$. Comparing \rf{4.53}
to the ordinary Schr\"odinger equation on $\cv_{\rm QM}$,
$i\hbar\, \partial_t|\psi\ra =\hat H |\psi\ra$,
this means that we may identify $|\psi\ra \equiv |\psi\ra_{\phi_0}$
and
\bel{4.54}
\hat H \,\equiv\, \co_H(\phi_0)
\ee
Eq.\rf{4.54} is a consistent generalization of the naive
''$H(\phi)$ becomes $H(\hat\varphi)$''--rule appropriate
for general phase--spaces. In the next section we shall
see that indeed, for a flat phase--space, $\co_H(\phi_0^a=0)=H(\hat\varphi)$.

An analogous discussion applies to arbitrary $(r,s)$--multispinors.
For an operator $\co_\rho$, for instance, eq.\rf{4.43} reads
\bel{4.55}
i\hbar\,\partial_t \co_{\rho(t)}(\phi)=
\left[\co_H(\phi), \co_{\rho(t)}(\phi)\right]
\ee
This is a local version of von Neumann's equation.
At $\phi=\phi_0$ it coincides with the usual  von Neumann
equation for the ordinary statistical operator 
$\co_\rho(\phi_0)\equiv \hat\rho$.
We are using the Schr\"odinger picture here, i.e., the entire
time dependence resides in the state vectors (pure states)
or in the statistical operator (mixed states). The operators
$\co_f(\phi)$ representing observables are time independent.
\vspace{3mm}

Up to now our discussion of the  spin bundles mostly
dealt with their local properties. Typically, in order to 
guarantee their existence globally, certain topological quantization
conditions must be fulfilled. Let $\{ \cp_\alpha\}$
be a contractible open covering of phase--space,
$\dis\cm_{2N}=\ccup{\alpha} \cp_\alpha$,
and let $\theta_{(\alpha)}$, $\Gamma_{(\alpha)}$ and
$\psi_{(\alpha)}$ locally represent a symplectic potential,
a \cg--valued connection and an arbitrary section through
the spin bundle on the patch $\cp_\alpha$. On the 
intersections $\cp_\alpha \cap \cp_\beta$, the symplectic
potentials are related by an abelian gauge transformation,
$\theta_{(\alpha)}=\theta_{(\beta)} +d\lambda_{\alpha\beta}$
and $\Gamma_{(\alpha)}$ is obtained from $\Gamma_{(\beta)}$ by
a Yang--Mills gauge transformation \rf{3.7} with a $G$--valued transition
function $U\equiv U_{\alpha\beta}$.
Likewise, $\psi_{(\alpha)}=t_{\alpha\beta} U_{\alpha\beta}\psi_{(\beta)}$
with $t_{\alpha\beta}\equiv \exp(-i\lambda_{\alpha\beta}/\hbar)$.

When we parallel transport a spinor from a point in $\cp_\alpha$
to a point in $\cp_\beta$, eq.\rf{4.40} is to be 
understood in the sense that the transition functions
$t_{\alpha\beta}$ and $U_{\alpha\beta}$ are applied to the
spinor at an arbitrary point in $\cp_\alpha\cap\cp_\beta$
which is visited by the curve \cc.

If we ignore the transition functions $U_{\alpha\beta}$ for a 
moment (i.e., assume that $U_{\alpha\beta}=1$ for all $\alpha,\beta)$,
the 1--forms $\theta_{(\alpha)}$ and the parameters $\lambda_{\alpha\beta}$
define a line bundle over $\cm_{2N}$ with curvature $\omega$. 
This is precisely the Kostant--Souriau prequantum bundle \cite{wood}
which is familiar from geometric quantization. A simple Dirac--type
argument shows that this bundle exists provided 
$\int_\Sigma \omega/(2\pi\hbar)$
is integer for any 2--cycle $\Sigma$, i.e., if 
$[\omega/(2\pi\hbar)]\in H^2(\cm_{2N}, Z)$.
Inequivalent bundles are classified by the cohomology class
of the 2--form $c_1=\omega/(2\pi\hbar)$, the first Chern class.
It would be interesting to have an analogous classification
of the bundles with structure group $G$, i.e., with the 
$U_{\alpha\beta}$'s retained. No general results are available
in this situation, but in ref.\cite{frad} the important
special case when the structure group can be reduced to
$[\Sp \times {\it U}(1)]/Z_2 $ has been discussed. One
finds that the above quantization condition is modified 
by the ''metaplectic anomaly'' involving the curvature
of the bundle of unitary frames. In ref.\cite{metone} 
we investigated a similar anomaly within our present
framework (at the semiclassical level). It can be used
to detect topological obstructions for the existence
of a metaplectic spin structure on $\cm_{2N}$ and is
closely related to the Maslov index.
 
\mysection{Flat Phase--Space}
In this section we
continue our investigation for the special case
of a flat phase--space. Besides being very important
in its own right, it illustrates the general ideas of
the previous sections in a rather explicit form.

\subsection{$\tilde\Gamma_a$ from the path--integral}
From now on we identify phase--space with the
symplectic plane $(\R^{2N}, \omega)$. We use global
Darboux coordinates $\phi^a=(p^k,q^k)$ with respect to which
$\omega_{ab}$ assumes the standard form \rf{2.3}. As before,
we employ the notation $\hat\varphi^a=(\hat\pi^k,\hat{x}^k)$
for the Heisenberg operators in the fiber. For the
symplectic connection we choose the trivial one,
$\Gamma_{abc}=0$. Hence $R_{abcd}=0$ and eq.\rf{4.27}
is solved by $r_a=0$ therefore. Thus the connection \rf{4.28}
boils down to
\bel{5.1}
\tilde\Gamma_a=\omega_{ab}\,\hat\varphi^b
\ee
This connection is independent of $\phi$. Its curvature
$\omega_{ab}$ is entirely due to the commutator terms in
$\Omega_{ab}$, which shows quite clearly the non--abelian
nature of the underlying gauge group. The most general
admissible connection for flat space was given in \rf{4.7}.
Fedosov's subsidiary condition (see the Appendix)
amounts to the choice $\alpha_a=0$, which we shall adopt
henceforth. The covariant derivative
\bel{5.2}
\frac{D}{D\phi^a}\equiv D_a = \partial_a
+\tfrac{i}{\hbar}\,\omega_{ab}\,\hat\varphi^b
\ee
decomposes according to
\ba\label{5.3}
\frac{D}{Dq^k} 
&=& \frac{\partial}{\partial q^k}-\frac{i}{\hbar}\hat\pi^k
=\frac{\partial}{\partial q^k}-\frac{\partial}{\partial x^k}
\\[.3cm]
\label{5.4}
\frac{D}{Dp^k} 
&=& \frac{\partial}{\partial p^k}+\frac{i}{\hbar}\hat{x}^k
=\frac{\partial}{\partial p^k}+\frac{i}{\hbar} x^k
\ea
The second equality in eqs.\rf{5.3} and \rf{5.4}, respectively,
holds in the representation in which $\hat{x}^k$ is diagonal.

Let us return to the path--integral of section 2. In eq.\rf{2.29} 
we wrote down an {\it exact} formula which relates two different
path--integrals which differ by a shift of the variable of
integration. If it was not for the nontrivial boundary conditions,
such a shift of a dummy variable would not have any effect.
The exponential in eq.\rf{2.29} is a consequence of these
nontrivial boundary conditions. The matrix element 
$\la q_2, t_2 | q_1, t_1\ra_H$ 
is manifestly independent of the
shift, $\Phi_{\rm cl}(t)$. Hence the $\Phi_{\rm cl}$--dependence of
$\la x_2, t_2 | x_1, t_1\ra_{\cal H}$ 
cancels precisely that of the exponential. 

Let us consider
eq.\rf{2.29} for two different trajectories,
$\Phi_{\rm cl}(t)$ and $\Phi_{\rm cl}(t)+\delta\phi(t)$.
Both of them are classical solutions, and 
$\delta\phi\equiv (\delta p_{\rm cl}, \delta q_{\rm cl})$
solves Jacobi's equation \rf{2.24} therefore. We would
like to eliminate
$\la q_2, t_2 | q_1, t_1\ra_H$ 
by subtracting these two equations. This makes it necessary to 
combine the change of the trajectory with a corresponding
change of the arguments of the matrix element:
$\delta x_{1,2}=-\delta q(t_{1,2})$.
Then, by eq.\rf{2.32}, both equations refer to the same
values of $q_1$ and $q_2$, and their difference reads
\bel{5.5}
\begin{array}{l}
\la x_2-\delta q_{\rm cl}(t_2), t_2 | x_1-\delta q_{\rm cl}(t_1),t_1
\ra_{{\cal H}(\Phi_{\rm cl}+\delta \phi)}
\,-\,\la x_2, t_2 | x_1,t_1
\ra_{{\cal H}(\Phi_{\rm cl})}\qquad\qquad \\[.3cm]
\qquad\qquad \dis=
-\tfrac{i}{\hbar}\delta\left[S_\rcl +p_\rcl(t_2)\, x_2 
-p_\rcl(t_1)\, x_1 \right] \la x_2, t_2|x_1,t_1\ra_{{\cal H}(\Phi_\rcl)} 
\\[.3cm]
\dis\qquad\qquad =
-\tfrac{i}{\hbar}\left[x_2\,\delta p_\rcl(t_2) -x_1\, 
\delta p_\rcl(t_1)\right] 
\la x_2, t_2|x_1,t_1\ra_{{\cal H}(\Phi_\rcl)} 
\eay
\ee
In the last line of \rf{5.5} we used \rf{2.32} and 
$\delta S_\rcl=p_\rcl(t_2)\delta q_{\rm cl}(t_2) -p_\rcl(t_1) 
\delta q_{\rm cl}(t_1)$.
The variation of the matrix element due to the change of the 
trajectory alone, i.e., at fixed $x_{1,2}$, is given by
\bel{5.6}
\kern -2ex
\bay
\delta\la x_2, t_2 | x_1,t_1\ra_{\cal H}
&=&\dis\left[\delta q_\rcl(t_2)\frac{\partial}{\partial x_2}
 +\delta q_\rcl(t_1) \frac{\partial}{\partial x_1}
-\tfrac{i}{\hbar}
\left(x_2\, \delta p_\rcl (t_2) - x_1\, \delta p_\rcl(t_1)\right)
\right]
\\
&&\cdot
\la x_2, t_2 | x_1,t_1\ra_{\cal H}
\eay
\ee
This is an exact result; no semiclassical expansion has been
invoked. 

There exists an alternative derivation of eq.\rf{5.6}
which makes no reference to the initial path--integral before 
the shift. If one varies the classical background trajectory
directly in the path--integral \rf{2.30}, performs a 
compensating transformation on the integration variables
and uses Jacobi's equation for $\delta\phi$ in order to eliminate
the terms linear in $\varphi^a$, one reproduces precisely eq.\rf{5.6}.

If one regards the matrix element as a function of the
variables referring to the final point, eq.\rf{5.6} shows that
\ba\label{5.7}
\left[\frac{\partial}{\partial q^k(t_2)} -\frac{\partial}{\partial x^k_2}
\right]
\la x_2, t_2 | x_1,t_1\ra_{\cal H}
&=&0\\[.3cm]
\label{5.8}
\left[\frac{\partial}{\partial p^k(t_2)} +\frac{i}{\hbar} x^k_2
\right]
\la x_2, t_2 | x_1,t_1\ra_{\cal H}&=&0
\ea
The interpretation of these equations is as follows. If we allow
$\delta\phi(t)$ to run through a complete set of Jacobi fields,
the trajectories $\Phi_\rcl(t) +\delta\phi(t)$
connect an infinitesimal neighborhood of $q_1$ to a 
corresponding neighborhood of $q_2$. Fixing
$\delta p_\rcl(t_2)$ and $\delta q_\rcl(t_2)$ means picking
a specific trajectory. Comparing eqs.\rf{5.7} and \rf{5.8}
to eqs.\rf{5.3} and \rf{5.4} we see that the transition
matrix element responds to a change of the classical
background trajectory in precisely such a way that
$\psi^x(p(t_2), q(t_2))\equiv \la x, t_2 | x_1,t_1\ra_{\cal H}$
is covariantly constant with respect to the abelian connection.

This calculation provides us with an independent derivation
of the connection $\tilde\Gamma_a$ and makes it explicit
that conventional quantum mechanics ''knows'' which connection
we have to use in the Hilbert bundle approach.

\subsection{The operators $\co_f$ and Moyal's equation}
On a flat phase--space the covariant derivative for operators
reads $D_a=\partial_a-\hat\partial_a$ and the
corresponding exterior differential for operator--valued
differential forms is $D=d-\delta$. The covariantly
constant fields $\co_f$ satisfy
\bel{5.9}
\partial_a\co_f +\tfrac{i}{\hbar}\,\omega_{ab}
\left[\hat\varphi^b,\co_f\right]=0
\ee
A simple calculation or eq.\rf{a.35} from the Appendix
show that for every analytic $f$ eq.\rf{5.9} is solved by
\bel{5.10}
\co_f(\phi)=f(\phi+\hat\varphi)
\ee
This operator is understood to be Weyl--ordered, i.e.,
it is defined by the following expansion in terms of
symmetrized operator products:
\bel{5.11}
\bay
\co_f(\phi) &=& \dis \sum_{m=0}^\infty 
\frac{1}{m!} \,
\partial_{a_1}\cdots\partial_{a_m}f(\phi)\,
\hat\varphi^{(a_1} \cdots \hat\varphi^{a_m)}
\\
&\equiv& \dis \exp\left(\hat\varphi^a\partial_a\right)f(\phi)
\eay
\ee
As it should be, the lowest order term of this power series
is the classical observable $f(\phi)$.

Let us evaluate the product of two $\co_f$'s. With the
notation $\partial_a^{(1,2)}\equiv \partial/\partial\phi^a_{1,2}$
we have 
\bel{5.12}
\kern -1.5ex
\bay
\co_f(\phi)\co_g(\phi) 
&=&
\dis
\exp\left(\hat\varphi^a\partial_a^{(1)}\right)
\exp\left(\hat\varphi^b\partial_b^{(2)}\right)
f(\phi_1)g(\phi_2)\Big\vert_{\phi_{1,2}=\phi}
\\[.3cm]
&=&
\dis
\exp\left[\hat\varphi^a\left(\partial_a^{(1)}+\partial_a^{(2)}\right)\right]
\exp\left[i\tfrac{\hbar}{2}\partial_a^{(1)}\omega^{ab}\partial_b^{(2)}\right]
f(\phi_1)g(\phi_2)\Big\vert_{\phi_{1,2}=\phi}
\eay
\ee
where the Baker--Campbell--Hausdorff formula was used. In order
to evaluate the coincidence limit we employ the identity
\bel{5.13}
G(\partial_a^{(1)}+\partial_a^{(2)})\,
F(\phi_1,\phi_2)\Big\vert_{\phi_{1,2}=\phi}
=G(\partial_a)\,F(\phi,\phi)
\ee
which holds true for arbitrary  analytic functions $F$ and $G$.
Hence
\bel{5.14}
\co_f(\phi)\,\co_g(\phi)=
\exp\left[\hat\varphi^a\partial_a\right](f*g)(\phi)
\ee
or 
\bel{5.15}
\co_f(\phi)\,\co_g(\phi)=\co_{f*g}(\phi)
\ee
with the product
\bel{5.16}
(f*g)(\phi)\equiv 
f(\phi)\exp\left[i\tfrac{\hbar}{2}
\stackrel{\leftarrow}{\partial}_a
\omega^{ab}
\stackrel{\rightarrow}{\partial}_b
\right]g(\phi)
\ee
As a consequence of eq.\rf{5.15}, the quantum
observables form a closed commutator algebra
\bel{5.16a}
\left[\co_f(\phi), \co_g(\phi)\right]
=i\hbar\,\co_{\{f,g\}_M}(\phi)
\ee
with the structure constants given by the Moyal bracket
$\{f,g\}_M\equiv (f*g-g*f)/i\hbar=\{f,g\}+O(\hbar^2)$.

Thus we have made contact with the familiar phase--space
formulation of quantum mechanics:
\rf{5.16} is precisely the ''star product'' appearing in 
the Weyl--Wigner--Moyal symbol calculus \cite{flato,ber}
which underlies the deformation theory approach to quantization.
The star product is an associative but noncommutative product on
the space of smooth functions over phase--space. It is the image
of the operator  product under the so--called symbol
map which establishes a one--to--one correspondence
between operators on $\cv_{\rm QM}$ and $c$--number functions
over $\cm_{2N}$.

The star product \rf{5.16} is appropriate if $\cm_{2N}$ is a flat
symplectic space. It is an important question if and how it can
be generalized to an arbitrary symplectic manifold \cite{bor}.
Fedosov \cite{fed} has shown that a star product indeed exists 
for any symplectic manifold and he devised an iterative
method to determine it. Within our present framework, his 
idea is as follows. Given \mbox{$\co_{f*g}$}, one can recover
\mbox{$f*g$} by extracting the term proportional to the unit operator.
Therefore, for an arbitrary curved phase--space, eq.\rf{5.15}
can be used as the {\it definition} of a star product. Usually no
closed--form solution for the $\co_f$'s is available then, but
it is always possible to insert the 
series expansion \rf{4.31} on the LHS of \rf{5.15}
and to construct the star product iteratively.
As this method has already been discussed in 
refs.\cite{fedbook, fed} we shall not go into any
details here. (See also refs.\cite{emm,quan}.)

In the Appendix we use a symbol calculus in order
to represent operators $A(\phi,\hat\varphi)$
in terms of symbols $A(\phi,y)$. Here $y$ is a coordinate
on a flat auxiliary phase--space which can be identified
with the tangent space at $\phi$ equipped with the constant
symplectic structure. Accidentally, the star product
\rf{5.16} for flat space has the same structure as the
$\circ$--product \rf{a.2}. For a general $\cm_{2N}$ the
$*$--product becomes more complicated, but as the
$\circ$--product refers to the flat tangent spaces
$(T_\phi\cm_{2N},\omega)$ it is always given by 
\rf{a.2}.

In section 4.3 we mentioned that the dynamics 
of mixed states is governed by the local von Neumann
equation \rf{4.55}. A priori this is an operatorial 
equation for the generalized statistical operator
$\co_\rho(\phi)$, but it can be converted to an equivalent
$c$--number equation for the function $\rho(\phi;t)$.
If we use \rf{5.16a} in \rf{4.55} and project on the
term proportional to the unit operator it follows that
\bel{5.17}
\partial_t\rho(\phi;t)=\{H,\rho(t)\}_M
\ee
Conversely, \rf{5.17} implies \rf{4.55} because both 
$\co_H$ and $\co_\rho$ are of the type \rf{5.10}.

Remarkably enough, eq.\rf{5.17} is precisely Moyal's
equation \cite{lj} for the time evolution of the 
pseudodensity $\rho(\phi;t)$, i.e., for the Weyl symbol
of the {\it ordinary} statistical operator $\hat\rho$ on
$\cv_{\rm QM}$. This fact can be understood as follows.
The fiberwise symbol calculus of the appendix
represents operators acting on $\cv_\phi$ in terms of
$c$--number functions over the flat ''auxiliary phase--space''
$(T_\phi\cm_{2N},\omega)$.
This construction applies in particular at the reference
point $\phi=\phi_0$ so that operators on $\cv_{\rm QM}$ are
changed into functions over the tangent space of $\phi_0$.
As we are dealing with a flat phase--space here, we may
identify $T_{\phi_0}\cm_{2N}$ with the manifold 
$\cm_{2N}=\R^{2N}$ itself and regard symbols such as
\mbox{[symb($\hat\varphi^a)](\phi)=\phi^a$} as functions 
$\cm_{2N}\to\R$. As a result, we recover exactly
the traditional symbol calculus which relates operators
on $\cv_{\rm QM}$ to functions over the classical
phase--space. For the statistical operator 
$\hat\rho=\co_\rho(\phi_0)$ we have in particular
\bel{5.18}
\left[ {\rm symb}\{\hat\rho\}\right](\phi)
=\left[ {\rm symb}\{\rho(\phi_0+\hat\varphi)\}\right](\phi)
=\rho(\phi_0+\phi)
\ee
It is convenient to identify the reference point with
the origin $\phi_0^a=0$. In this case the Weyl symbol
of the ordinary statistical operator equals
precisely the function $\rho$ which characterizes the
local operator $\co_\rho(\phi)$, and eq.\rf{5.17} has
its usual meaning therefore.

With the choice $\phi^a_0=0$ the conventional quantum
observables are
\bel{5.18a}
\hat{f}\equiv \co_f(0)=f(\hat\varphi)
\ee
They are obtained from the classical observables $f(\phi)$
by the usual substitution $\phi\to\hat\varphi$ together
with the Weyl ordering prescription.

\subsection{The parallel transport operators}
For the connection \rf{5.1}, the parallel transport
operator $V(s)\equiv V[\cc(\phi(s), \phi_0)]$
of \rf{3.48} satisfies the differential equation
\bel{5.19}
i\hbar\,\frac{d}{ds} V(s) =\dot\phi^a(s)\,\omega_{ab}\,
\hat\varphi^b \,V(s)
\ee
It can be solved explicitly in terms of the Weyl operators \cite{lj}
\bel{5.20}
T(\phi)\equiv \exp\left(\tfrac{i}{\hbar}\,\phi^a\,\omega_{ab}
\,\hat\varphi^b\right)
\ee
They provide a projective representation of the translations on the
symplectic plane. We shall need the following properties:
\bel{5.21}
T(\phi)^\da\hat\varphi^aT(\phi)=\hat\varphi^a+\phi^a
\ee
\bel{5.22}
T(\phi)^\da=T(\phi)^{-1}=T(-\phi)
\ee
\bel{5.23}
T(\phi_1)T(\phi_2)=\exp\left[\tfrac{i}{2\hbar}\phi^a_1\omega_{ab}\phi^b_2
\right] T(\phi_1+\phi_2)
\ee
\bel{5.24}
\bay
\partial_aT(\phi)&=&\dis \tfrac{i}{\hbar}\omega_{ab}
\left[\hat\varphi^b-\tfrac{1}{2}\phi^b\right] T(\phi)
\\
&=&\dis \tfrac{i}{\hbar} \omega_{ab} T(\phi)
\left[\hat\varphi^b+\tfrac{1}{2}\phi^b\right] 
\eay
\ee 
Using these formulae it is easy to verify that
the solution of \rf{5.19} with $V(0)=1$ is given by
\bel{5.25}
V(s)=\exp\left[
 \tfrac{i}{2\hbar}\int_0^s ds^\prime \, \dot\phi^a(s^\prime)
\,\omega_{ab}\,\phi^b(s^\prime)\right]\,
T\left(\phi(s)\right)^\da \,T\left(\phi(0)\right)
\ee
Setting $s=1$ and $\phi(0)=\phi_0$, $\phi(1)=\phi$, eq.\rf{5.25}
is of the form \rf{4.37} with the path--independent part
\bel{5.26}
\tau(\phi, \phi_0) =T(\phi)^\da\, T(\phi_0)
\ee
The path--dependent phase factor
in eq.\rf{4.37} contains the symplectic potential \rf{4.13}.
Incidentally, if we solve the parallel transport equation
for the flat connection \rf{4.14} rather than the abelian
one with curvature $\omega$, this phase factor is absent
and one has $V[\cc(\phi,\phi_0)]=T(\phi)^\da T(\phi_0)$.

For convenience we identify $\cv_{\rm QM}$ with the Hilbert
space at the origin of the symplectic plane, $\phi_0=0$,
and we set $\tau(\phi)\equiv \tau(\phi,0)$.
We pick some conventional quantum mechanical wave function
$\Psi(x)$ in $\cv_{\rm QM}$ and interpret it as a spinor
in the fiber at $\phi=0$: $\psi(0)^x\equiv \Psi(x)$.
This wave function can be parallel transported to any point
of phase--space: 
$\psi_{\cal C}(\phi)^x =V[\cc(\phi,0)]^x_{~y} \psi(0)^y$.
Using the position--space matrix elements of the Weyl--operators
\cite{lj} one finds for $\phi\equiv (p,q)$:
\bel{5.27}
\tau(p,q)^x_{~y} \,\psi(0)^y=
\exp\left[-\tfrac{i}{\hbar}p\left(x+\tfrac{1}{2}q\right)\right]
\psi(0)^{x+q}
\ee
As we mentioned already in section 4.3, any vector in the
reference Hilbert space can be consistently extended to 
a spinor field on an arbitrary Lagrangian submanifold $\ck$
which contains $\phi_0$.
The most important example of such a submanifold is the
configuration space $\ck=\{(p=0,q)|q\in \R^N\}$.
For every path $\cc$ which connects $\phi_0=0$ to the point
$\phi=(0,q)$ and stays in the plane $p=0$ one has 
$\int_{\cal C}\theta=0$. 
Hence, when transported to $\phi$, the wave function reads
\bel{5.28}
\psi(0,q)^x=\Psi(q+x)
\ee
This defines a covariantly constant spinor field on $\ck$.
In this simple example the meaning of the background--quantum
split symmetry is particularly clear.
The field \rf{5.28} depends only on the invariant combination
$q+x$, which is annihilated by \rf{5.3}. In the Hilbert space
located at $(0,q)$, the ''quantum'' part of the argument of
$\Psi,$ $x$, is augmented by the ''background'' piece $q$.
But also the observables to be used at $(0,q)$ are different,
$\co_f(0,q)=f(\hat\pi,q+\hat{x})$,
and it does not matter therefore in which Hilbert space
expectation values or matrix elements are calculated.

\subsection{Time evolution and bilinear covariants}
In order to write down the universal equation of motion
\rf{4.43}, \rf{4.44} one has to know the generator $\co_H$
of local frame rotations. On flat phase--space we have
the closed--form solution
\bel{5.29}
\co_H(\phi)=H(\phi+\hat\varphi)
\ee
In order to describe the relation of our approach to 
the path--integral quantization, it is helpful to pull out
the first two terms of its power series in $\hat\varphi$:
\bel{5.30}
\ch(\hat\varphi;\phi) \equiv H(\phi+\hat\varphi)
-\partial_a H(\phi)\,\hat\varphi^a-H(\phi)
\ee
Now we insert \rf{5.29} with \rf{5.30} into \rf{4.44}
and rewrite the terms linear in $\hat\varphi$ by a trick
similar to the one which we used in section 4.3 in the 
investigation of the semiclassical limit. For a single
upper spinor index, say,
\bel{5.31}
\partial_aH(\phi) \,\hat\varphi^a
=i\hbar\left(h^aD_a - h^a\partial_a\right)
\ee
As a result, eq.\rf{4.44} assumes the following 
suggestive form:
\bel{5.32}
\bay
\cl_H(\phi) \,
\chi(\phi)^{x_1\cdots x_r}_{y_1 \cdots y_s}
&=&\dis
\left[h^a \partial_a -h^a D_a +\tfrac{i}{\hbar}(r-s)H\right]
\chi^{x_1\cdots x_r}_{y_1 \cdots y_s}
\\[.3cm]
&& \dis
+\tfrac{i}{\hbar}\, \ch (\hat\varphi ; \phi)^{x_1}_{~z} 
\,\chi^{z\cdots}_{y_1\cdots}
-\tfrac{i}{\hbar}\, \chi^{x_1\cdots}_{z\cdots}
\,\ch (\hat\varphi ;\phi)^z_{~y_1}\, \pm \cdots
\eay
\ee
While this equation is exact, its structure is similar
to the semiclassical $\cl_H^{\rm scl}$ of \rf{4.47}.
In fact, for the special case of $\cm_{2N}$ flat,
eq.\rf{5.32} can be thought of as the correct all--order
generalization of the metaplectic ''Lie--derivative''.
In general, the operator
\bel{5.33}
\ch(\hat\varphi ; \phi) =
\sum_{m=2}^\infty \frac{1}{m!}\,
\partial_{a_1}\cdots\partial_{a_m} H(\phi)\,
\hat\varphi^{(a_1}\cdots \hat\varphi^{a_m)}
\ee
contains terms of arbitrarily high degree. Only in the 
semiclassical limit it becomes an element of $\mpl$,
\bel{5.34}
\ch=\tfrac{\hbar}{2}\,\partial_a\partial_b H(\phi) \Sigma^{ab} +O(3)
\ee
and \rf{5.32} reduces to \rf{4.47} then. Applying $\cl_H$ to
covariantly constant fields so that $D_a\chi=0$ and 
disregarding the term proportional to $(r-s)$ which produces
a phase only, eq.\rf{5.32} is reminiscent of a Lie--derivative.
However, we emphasize that
$\cl_H(\phi)$ generates only a local frame rotation but no
diffeomorphism on $\cm_{2N}$. The $h^a\partial_a$--term in 
$\cl_H(\phi)$ originates from the condition $D_a\chi =0$ which
allows us to trade shifts on the base manifold for transformations
in the fibers.

In order to better understand the properties of 
$\cl_H$ and $\ell_H$ it is quite instructive to look at
the following bilinears formed from two arbitrary spinors
$\psi$ and $\chi$:
\bel{5.35}
\bay
E(\phi) &=&{}_\phi \la\chi|\psi\ra_\phi \\
T^a(\phi) &=&{}_\phi \la\chi|\,\gamma^a\,|\psi\ra_\phi \\
R^{ab}(\phi) &=&{}_\phi \la\chi|\,\Sigma^{ab}\,|\psi\ra_\phi 
\eay
\ee
These bilinears are related to the generators of the group
${\it I Mp}(2N)$, the semidirect product of $\Mp$ with the Weyl group
\cite{metino}.
Their metaplectic Lie--derivative \rf{2.13} is easily worked 
out \cite{metino}:
\ba
\ell_H E & = & h^a\partial_a E \label{5.36} \\
\ell_H T^a & = & h^b\partial_b T^a-\partial_bh^a T^b \label{5.37} \\
\ell_H R^{ab} & = & h^c\partial_c R^{ab}
                   -\partial_c h^aR^{cb}-\partial_c h^b R^{ac}
 \label{5.38} 
\ea
Obviously, $E$, $T^a$ and $R^{ab}$ transform as a scalar, a vector
and a symmetric tensor, respectively. On the other hand, applying
$\cl_H$ yields\footnote{ Here we generalize the definition of $\cl_H$ to
tensorial quantities with coordinate or frame indices in such a 
way that $\cl_H\gamma^{ax}_{~~y}=0$. For a $c$--number vector
$v^a(\phi)$, say, one needs the nonlinear transformation
$\cl_H v^a(\phi)=-h^a(\phi+v)$. This is precisely as it should be,
because we may identify the flat phase space with the
''reference tangent space'' at $\phi=0$. Therefore $v^a\equiv v^a(\phi=0)$
can be identified with the coordinate $\phi^a$ on $\cm_{2N}$.
Hence the above nonlinear transformation yields the equation of
motion $\partial_t v^a = h^a(v)$, which is nothing but Hamilton's
equation for $v^a\equiv \phi^a$.}
\ba
\cl_H \,E(\phi) &=& 0 \label{5.39} \\
\begin{array}{c}
\cl_H \,T^a(\phi) 
\\
\qquad
\end{array}
&
\begin{array}{c}=\\=\end{array}
& 
\begin{array}{l}
-\sqrt{2/\hbar}\,\,
{}_\phi\la\chi| \,h^a(\phi+\hat\varphi)\,|\psi\ra_\phi
\\
-\sqrt{2/\hbar}\, h^a E -\partial_b h^a \,T^b +O(2)
\end{array}
\label{5.40}\\[.3cm]
\begin{array}{c}
\cl_H\, R^{ab}(\phi) 
\\
\qquad
\end{array}
&
\begin{array}{c}=\\=\end{array}
& 
\begin{array}{l}
-\hbar^{-1}\,\, 
{}_\phi\la\chi|\, \hat\varphi^{(a} h^{b)}(\phi+\hat\varphi)\,
+\, h^{(a}(\phi+\hat\varphi)\hat\varphi^{b)}\,
|\psi\ra_\phi
\\
-\sqrt{2/\hbar}\, h^{(a}T^{b)}
-\partial_ch^a\, R^{cb} -\partial_ch^b\, R^{ac} +O(3)
\end{array}\label{5.41}
\ea
Several lessons can be learned from these results.
First, $\ell_H$ contains the generators of $\Mp$ which
are quadratic in $\hat\varphi$. Therefore $\ell_H$ preserves
the number of $\gamma$--matrices and generates a linear transformation
of the bilinears. This is not the case for $\cl_H$ which is related
to the $W_\infty$--type Lie algebra $\cg$ spanned by all
monomials of the form $\hat\varphi^{(a_1}\cdots \hat\varphi^{a_n)}$.
If one expands $h(\phi+\hat\varphi)$ in eq.\rf{5.40} one obtains
terms of arbitrarily high degree. Second, we observe that for
generic spinor fields $\psi$ and $\chi$ even the terms in $\cl_H$
of lowest degree  do not agree with $\ell_H$. 
In fact, $\cl_H=\ell_H+O(1)$ is expected to hold only
if the dynamical phase cancels (which is the case here)
and if the fields involved are covariantly constant.
Let us assume therefore that $|\psi\ra_\phi$ and ${}_\phi\la\chi|$
are obtained by parallel transporting $|\psi\ra_0$
and ${}_0\la\chi|$ from the reference point $\phi=0$ to $\phi$.
The path--dependent phase cancels from the bilinears, and using
$\tau(\phi,0)=T(-\phi)$ in \rf{4.40} one finds that
their functional form is highly constrained:
\bel{5.42}
\bay
E(\phi)&=& E(0) \\
T^a(\phi) &=& T^a(0)-\sqrt{2/\hbar}\, \phi^a\, E(0)\\
R^{ab}(\phi) &=& R^{ab}(0) 
-\sqrt{2/\hbar} \,\phi^{(a}T^{b)}(0) +\hbar^{-1}\,\phi^a\phi^b\,E(0)
\eay
\ee
Using \rf{5.42} in  the semiclassical expansions of
\rf{5.39}--\rf{5.41} it is easy to check that
the terms of nonpositive degree generated by $\cl_H$
coincide precisely with those of $\ell_H$, 
eqs.\rf{5.36}--\rf{5.38}.
For instance, the fact that $E$ is strictly constant
for covariantly constant spinors reconciles \rf{5.36}
with \rf{5.39}.
\vspace{3mm}

To close with, let us look at the all--order
dynamics of the world--line spinors. They provide a natural
link between our approach and standard path--integral
quantization also beyond the semiclassical limit. 
We shall make contact with the {\it exact} shifted 
path--integral \rf{2.30}. To this end we consider 
an arbitrary classical trajectory $\Phi_\rcl(t)$ 
which, at time $t=0$, starts at the point $\phi_0$.
We identify the fiber at the initial point with the
quantum mechanical Hilbert space, $\cv_{\phi_0}\cong \cv_{\rm QM}$.
Furthermore, we pick some time--dependent state vector
$\psi(\phi_0;t)$ in $\cv_{\rm QM}$; it evolves according
to $i\hbar \,\partial_t \psi(\phi_0;t)=\co_H(\phi_0) \psi(\phi_0;t)$.
For every fixed
time $t$, we parallel transport this state along the classical
trajectory from $\phi_0$ to $\Phi_\rcl(t)$:
\bel{5.43}
\bay
\psi(\Phi_\rcl(t) ; t)
&=&\dis
V\left[\cc\left(\Phi_\rcl(t),\phi_0\right)\right]
\psi(\phi_0 ;t)
\\
&=&
V\left[\cc\left(\Phi_\rcl(t),\phi_0\right)\right]
U(\phi_0; t)\,
\psi(\phi_0 ;0)
\eay
\ee
In this manner we construct a spinor field along $\Phi_\rcl$:
$\eta(t) \equiv \psi\left(\Phi_\rcl(t) ;t\right) \in \cv_{\Phi_\rcl(t)}$.
Taking the time derivative of \rf{5.43} one obtains
\bel{5.44}
i\hbar\,\partial_t \eta =\left[\co_H(\Phi_\rcl)
-\partial_aH(\Phi_\rcl) \,\hat\varphi^a\right]\eta
\ee
where the term linear in $\hat\varphi$ results from 
differentiating the parallel transport operator. Thus,
since $H$ is constant along $\Phi_\rcl$,
\bel{5.45}
i\hbar \,\partial_t \eta =
\left[\ch(\hat\varphi ;\Phi_\rcl) +H(\phi_0)\right]\eta
\ee
Comparing this equation of motion to eq.\rf{2.33}
we see that, apart from a trivial phase, the world line
spinor $\eta$ is precisely the same thing as the matrix
element represented by the shifted path integral of
section 2:
\bel{5.46}
\eta^x(t) =\exp\left(-\tfrac{i}{\hbar}H(\phi_0)t\right)
\la x,t | x_0, 0\ra_{\cal H}
\ee

In establishing this result we have understood the
all--order generalization of the semiclassical
world--line spinors and their
relation to the background--quantum split symmetry.
Both in the construction above and in the path--integral
derivation of \rf{2.33}, a classical trajectory was 
introduced as  a ''background'', but nevertheless
we were always dealing with  full--fledged 
quantum mechanics.

\mysection{Discussion and Conclusion}
In this paper, we have uncovered a hidden
gauge theory structure of quantum mechanics
which is not visible in its conventional formulations.
When this structure is made manifest, one realizes
that the conceptual framework of quantum theory shows
a remarkable similarity with a gauge theory, in
particular with general relativity.

We constructed a Yang--Mills--type theory on the
phase--space of an arbitrary hamiltonian system.
A local Hilbert space was associated to each point of
phase--space, and ''matter fields'' were introduced
which assume values in these spaces. We referred to
them as ''metaplectic spinors'' because under $\Sp$
they transform in its spinor, or metaplectic,
representation. The underlying gauge group $G$
is infinite--dimensional though. It consists of
all unitary frame rotations in the fiber $\cv$.

We discussed in detail how conventional quantum
mechanics can be reformulated within this general
setting, and we showed that this reformulation
provides a new way of understanding the structure of 
quantum theory and 
the transition
from classical to quantum mechanics. It turned out
that the rules of canonical quantization can be replaced
by two new ''postulates'' with a much clearer group theoretical
and geometrical interpretation.

The first postulate is of a purely group theoretical
nature and contains the transition from classical
mechanics to semiclassical quantum mechanics. In the
classical context, we are dealing with tensor fields over
phase--space. Under local frame rotations they transform
in some product of the vector representation of $\Sp$.
The first postulate says that we have to replace
the vector representation of $\Sp$ by the spinor
representation of its double covering $\Mp$ and
to use multispinors rather than tensors for the
description of physical states and observables. We
motivated this rule by applying it to the classical
Jacobi fields; the resulting world--line spinors are
precisely the semiclassical wave functions.

For nonlinear systems we need a second rule which
tells us how to recover the exact quantum theory from its
semiclassical approximation. The second postulate is that
the multispinor 
fields must be covariantly constant (up to a phase 
possibly) with respect to an arbitrary abelian
$\cg$--valued connection $\tilde\Gamma$. This flatness
condition means that the states and operators related
to the different local 
Hilbert spaces can all be identified by virtue of the
parallel transport defined by $\tilde\Gamma$. In this
manner, the equivalence of the Hilbert bundle approach 
and the
standard one--Hilbert space formulation of
quantum mechanics is established.
\vspace{3mm}

It is one of our main results that the second postulate
is equivalent to a very simple but deep symmetry
principle: invariance under the background--quantum
split symmetry. We visualize the exact quantum theory 
as the result of consistently sewing together
an infinity of local quantum theories, one at each point
of phase--space. Classically, if a particle sits at the
point $\phi$, the value of the observable $f$ is $f(\phi)$.
We assume that the quantum correction to this value is
locally determined by the expectation value of a certain
operator $\Delta \co_f(\phi)$ with respect to a vector
in $\cv_\phi$:
$\la \hat{f}\ra\equiv f(\phi)+{}_\phi\la\psi|\Delta\co_f(\phi)|\psi\ra_\phi$.
When applied to each point $\phi$ separately, this prescription
raises the following consistency problem.
For $f(\phi)=\phi^a$ and flat space, say,
we would like to interpret
$\la\hat\phi^a\ra=\phi^a+{}_\phi\la\psi|\hat\varphi^a|\psi\ra_\phi$
as the exact quantum mechanical expectation value of 
the position in phase--space. Clearly this is possible only
if $\la\hat\phi^a\ra$ is independent of the point $\phi$.
Going to a new point $\bar\phi$, the new state $|\psi\ra_{\bar\phi}$
must be such that the resulting change of the
$\hat\varphi$--expectation value cancels
the difference $\phi^a-\bar\phi^a$. The postulate of the
background--quantum split symmetry means that, more generally,
the value of $\la\hat{f}\ra$ should be independent
of $\phi$, i.e., that all the local quantum theories
agree on the
expectation value of the quantum observables
$\co_f=f+\Delta \co_f$. The form of these operators is
fixed by the condition 
$\dis \lim_{\hbar\to 0} \co_f(\phi)=f(\phi)$
together with 
the split symmetry; it implies that they must be
covariantly constant with
respect to an abelian connection.

The physical motivation for the second postulate comes
from semiclassical considerations again. Even if not in 
practice, at least from a conceptional point of view,
our quantization program first constructs a semiclassical
approximation of quantum mechanics as a link between the
classical and the exact quantum theory.
Roughly  speaking, the idea is to recover 
the exact theory from the totality of the semiclassical
wave functions calculated for all classical trajectories.
Let us assume we know the solutions $\eta(t)$ of the
semiclassical Schr\"odinger equation \rf{2.23} or \rf{2.33}
for all classical paths $\Phi(t)$. Heuristically, the
wave function $\eta_1(t)$ belonging to some trajectory
$\Phi_1(t)$ provides a reasonable approximation to the
complete theory within a tubular neighborhood of
$\Phi_1(t)$. Within this neighborhood, the neglected
nonlinearities should be irrelevant. Let us suppose there
is a nearby classical path $\Phi_2(t)$ such that the
tubular neighborhood within which its wave function
$\eta_2(t)$ is valid overlaps with the one of 
$\Phi_1(t)$.
Thus there is a region in phase--space where both 
semiclassical expansions apply. The expectation
value of the ''position'', say, is given by
$\Phi_1^a(t)+\bar\eta_1(t)\hat\varphi^a\eta_1(t)$ 
according to
the first, and by $\Phi_2^a(t)+\bar\eta_2(t)\hat\varphi^a\eta_2(t)$
according to the second one. These values must coincide
if both expansions are approximations to the same
exact theory. 
If we look at this situation at a fixed instant of
time, we are back to the above picture of local quantum
theories at a point, and it is clear that the necessary
consistency requirement is precisely that $\Phi(t)$ and
$\eta(t)$ must be connected by the split symmetry.

Furthermore, from a dynamical point of view, the ''current''
$\bar\eta(t)\hat\varphi^a\eta(t)$
has the same properties as the classical Jacobi
field $c^a(t)$. This yields a simple interpretation of the
first postulate as well: we have to take the ''square
root'' of the Jacobi field, $c^a\to \bar\eta\hat\varphi^a\eta$.
This substitution converts the position of a classical
particle propagating near $\Phi(t)$, i.e., 
$\Phi^a(t)+c^a(t)$, to the corresponding expectation
value in semiclassical quantum mechanics.
By virtue of the split symmetry, the ''quantum
Jacobi fields'' $\eta(t)$ can be glued together 
consistently to yield  fields $\psi^x(\phi;t)$. 
Their value at an arbitrary reference point can
be identified with the conventional
wave function:
$\Psi(x;t) \equiv \psi^x(\phi_0;t)$.

It is amusing to compare the process of gluing
together local semiclassical expansions to the
transition from special to general relativity.
The tubular neighborhoods surrounding $\Phi(t)$
correspond to the freely falling ''Einstein
elevators'' within which the laws of
special relativistic physics 
are a good 
approximation. If one tries to consistently patch up
the observations made in different such elevators, not
necessarily close to each other, one needs a connection,
or a parallel transport, and thus starts feeling the
curvature of space--time. Special--relativistic physics
corresponds here to the local semiclassical approximation
of quantum mechanics which, too, can be formulated
without a connection. Both in general relativity and in
our Hilbert bundle approach the connection is the
essential tool for ''globalizing'' local physics. However,
because of the strong constraints coming from the split
symmetry, this connection is not dynamical in the case
of quantum mechanics.

In many of our derivations and ''thought experiments'',
classical solutions played a central r\^ole. We
emphasize, however, that in order to actually
quantize a given system along the lines proposed here
no knowledge of these solutions and their moduli space
is necessary. All one has to do is to find an abelian
connection and to construct covariantly constant
sections, which is similar to the method of ref. \cite{frad}. 
From a purely pragmatic 
point of view, the subdivision of the quantization process according
to the two postulates is unnecessary.

Our main emphasis was on gaining
a better understanding of what it means to 
``quantize'' a hamiltonian system.
While the familiar rules of canonical quantization
can be formulated quite easily, their origin is rather
obscure still. 
We replaced them by two postulates which in our
opinion are much more natural and easier to understand
intuitively. 

The first postulate involves nothing but
changing the representation of the
''Lorentz group'' appropriate for 
phase--space. Comparing this to the situation in
space--time, we are
perfectly familiar with the idea that besides the spin-1
representation (photons, gluons, ...) also the spin--$\tfrac{1}{2}$
representation (electrons, quarks, ...) of the Lorentz
group is realized in nature. Hence it appears 
quite natural that also at the more fundamental
level of quantum theory in general nature takes
advantage of spinor representations. Moreover, the second
postulate is formulated in the same language of classical
differential geometry as general relativity or
Yang--Mills theory. It is our hope, therefore, that the
approach proposed here will help for instance in understanding
the interrelation of gravity and quantum theory
on a geometric basis.
\vspace{3mm}

\noindent
Acknowledgements:

\noindent
I am indebted to Ennio Gozzi for numerous stimulating discussions
which triggered these investigations. I am also grateful to
M.\,Bordemann, O.\,Dayi, G.\,Mack, H.--J.\,Matschull, H.\,R\"omer
and S.\,Sorella for helpful conversations.

\newpage

\begin{appendix}
\section*{Appendix}
\newcommand{\sect}[1]{\setcounter{equation}{0}
  \renewcommand{\theequation}{#1.\arabic{equation}}
 }
\sect{A}                  

This appendix serves two purposes: First, we show
how the operators on $\cv_\phi$ 
are represented
in a {\it fiberwise} Weyl--symbol calculus, and second, we
follow Fedosov \cite{fed,fedbook} and use this method in order
to solve eq.\rf{4.27} for $r$ and to establish eq.\rf{4.31}.

For the time being we consider a single copy of the 
typical fiber $\cv$. We represent operators $\hat{f}:\, \cv\to\cv$
by their Weyl symbols \cite{ber} $f={\rm symb}(\hat{f})$. They
are functions of the auxiliary variable
$y\equiv(y^a)\in\R^{2N}$: $f(y)=[{\rm symb}(\hat{f})](y)$. The precise
definition of the symbol map can be found in refs.\cite{ber,lj}.
Here we only mention some properties which will 
be needed later on. Under the symbol map, the 
operator product is mapped onto the $\circ$--product,
\bel{a.1}
{\rm symb}(\hat{f}\,\hat{g}) = 
{\rm symb}(\hat{f})\circ{\rm symb}(\hat{g})
\ee
which reads explicitly
\bel{a.2}
(f\circ g)(y)=f(y) \exp\left[i\frac{\hbar}{2}
\stackrel{\leftarrow}{\frac{\partial}{\partial y^a}}
\omega^{ab}
\stackrel{\rightarrow}{\frac{\partial}{\partial y^b}}
\right] g(y)
\ee
This is precisely the familiar ''star product'' for
functions over the symplectic vector space $(\R^{2N},\omega)$.
This auxiliary phase--space should be carefully distinguished
from the actual phase--space of the physical system
under consideration, $\cm_{2N}$. (We reserve the
conventional notation
''$*$'' for the star--product referring to $\cm_{2N}$.)

The Weyl symbol has the property $[{\rm symb}(\hat\varphi^a)](y)=y^a$
from which it follows that
\bel{a.3}
\left[{\rm symb}(\hat\varphi^a\hat\varphi^b)\right](y)
=y^a\circ y^b =y^ay^b+i\,\tfrac{\hbar}{2}\,\omega^{ab}
\ee
Upon symmetrization, the $\omega^{ab}$--term vanishes, and $y^ay^b$
is found to be the symbol of $\hat\varphi^{(a}\hat\varphi^{b)}$.
This result generalizes to arbitrary symmetrized products of $\hat\varphi$'s:
\bel{a.4}
\left[
{\rm symb}\left( \hat\varphi^{(a_1}\cdots \hat\varphi^{a_n)}\right)
\right](y)
=y^{a_1}y^{a_2}\cdots y^{a_n}
\ee

Now we return to the bundle of local Hilbert spaces 
$\cv_\phi$ and represent the operators $A(\phi):\, \cv_\phi \to \cv_\phi$
at all points $\phi$ by their symbols. More generally, we consider
operator--valued $k$--forms of the type
\bel{a.5}
A_{kl}(\phi,\hat\varphi,d\phi)
=
A_{a_1\cdots a_l \, b_1\cdots b_k}(\phi)
\,\hat\varphi^{(a_1}\cdots\hat\varphi^{a_l)}
\,d\phi^{b_1}\cdots d\phi^{b_k}
\ee
Their Weyl symbols read
\bel{a.6}
A_{kl}(\phi,y,d\phi)
=
A_{a_1\cdots a_l \, b_1\cdots b_k}(\phi)
\, y^{a_1}\cdots y^{a_l}\,
d\phi^{b_1}\cdots d\phi^{b_k}
\ee
We define the following linear operations on the space
of symbols $\dis A\equiv\sum_{kl} A_{kl}$.
The projectors $P_0$ and $P_{00}$:
\ba
(P_0 A) (\phi,y,d\phi) & =&  A(\phi,0,d\phi) = A_0(\phi, d\phi) \label{a.7}\\
(P_{00} A) (\phi,y,d\phi) & =&\,\,A(\phi,0,0)\,\,
 = A_{00}(\phi) \label{a.8}
\ea
The number operators $\cn_{d\phi}$, $\cn_y$ and $\cn$:
\ba
\cn_{d\phi} &=& d\phi^a \,{\bf i} (\partial_a) \label{a.9} \\
\cn_{y}\,\,\, 
&=& dy^a \,\frac{\partial}{\partial y^a} \label{a.10} \\
\cn  \,\,\,\,\,   &=& \cn_{d\phi} +\cn_y \label{a.11}
\ea
The nilpotent maps $\delta$, $\delta^*$ and $\delta^{-1}$:
\ba
\delta  &=&d\phi^a \,\frac{\partial}{\partial y^a}  \label{a.12}\\[.3cm]
\delta^* &=&y^a \,{\bf i} (\partial_a)  \label{a.13}\\[.3cm]
\delta^{-1} &=&\delta^* \,\cn^{-1} (1-P_{00}) \label{a.14}
\ea
Note that \rf{a.12} is nothing but eqs.\rf{4.20},
\rf{4.21} in symbol language.
In eq.\rf{a.9}, ${\bf i}(\partial_a)$ denotes
the contraction with the vector field $\partial_a$. It satisfies
$[{\bf i}(\partial_a), d\phi^b]=\delta^b_a$, which shows that
$\cn_{d\phi} A_{kl} =k A_{kl}$.
Likewise, $\cn_y A_{kl}=l A_{kl}$. It is easy to see that
\bel{a.15}
[\delta, \cn] =0 \qquad, \qquad [\delta^*, \cn]=0
\ee
and 
\bel{a.16}
\bay
\delta \,P_{00} &=& P_{00}\,\delta =0 \\
\delta^*\, P_{00} &=& P_{00}\,\delta^* =0 
\eay
\ee
The operators $\delta$, $\delta^*$ and $\cn$ constitute a kind of
supersymmetry algebra:
\bel{a.17}
\begin{array}{c}
\delta^2=0 \,\,  , \,\,  (\delta^*)^2=0 \\
\delta\delta^* +\delta^*\delta =\cn
\eay
\ee
It follows from \rf{a.15}, \rf{a.16} and \rf{a.17} that
\bel{a.18}
\delta\delta^{-1} +\delta^{-1}\delta+P_{00}=1
\ee
and that $\delta^{-1}$ is nilpotent, too: $(\delta^{-1})^2=0$.
Eq.\rf{a.18} is an important identity. It implies that any
operator valued differential form $A$ admits the
''Hodge decomposition''
\bel{a.19}
A=A_{00}+\delta\delta^{-1}A+\delta^{-1}\delta A
\ee

This  decomposition is the main tool for solving
eq.\rf{4.27},
\bel{a.20}
\delta r = R +\nabla r +\tfrac{i}{\hbar} \,r^2
\quad , \quad R\equiv \Omega(\Gamma)
\ee
Here $r_a(\phi,y)$ are the symbols of the operators
$r_a(\phi,\hat\varphi)$; since the latter are Weyl
ordered by definition, their symbols are obtained 
by simply replacing $\hat\varphi^a\to y^a$
everywhere. Let us apply \rf{a.19} to $r\equiv r_a(\phi,y)d\phi^a$:
\bel{a.21}
r=r_{00} +\delta \delta^{-1} r +\delta^{-1}\delta r
\ee
Clearly, $r_{00}=0$ because $r$ is a 1--form. A priori
eq.\rf{a.20} admits many different solutions. For our program
of implementing the background--quantum split symmetry it
is necessary to know only one particular solution and not
all of them. In order to make the solution unique, one
may impose a subsidiary condition. A very convenient choice is
\bel{a.22}
\delta^{-1}r=0 \qquad {\rm or} \qquad y^a\,r_a(\phi,y)=0
\ee
Assuming, as always, an analytic dependence on $y^a$, this
implies that
\bel{a.23}
r_a(\phi,0)=0 \qquad {\rm and}
\qquad
\partial^y_{(b_1}\cdots \partial^y_{b_n}
\,r_{a)} (\phi,0)=0
\ee
This means that in particular $r$ contains no term
proportional to the unit operator. For this choice,
eq.\rf{a.21} becomes $r=\delta^{-1}\delta r$ or
\bel{a.24}
r=\delta^{-1} R +\delta^{-1} 
\left[\nabla r+\tfrac{i}{\hbar}\, r^2 \right]
\ee
The crucial observation is that the operator acting
on $r$ on the RHS of this equation increases the degree.
(The degree is defined as in section 4.2 with $y^a$ playing
the r\^ole of $\hat\varphi^a$ now, i.e., ${\rm deg}(y^a)=1$.)
In fact, $\delta^{-1}$ increases the degree by one unit, and
since ${\rm deg}(\Gamma_a)=2$ for a symplectic connection,
$\nabla$ preserves the degree. 
This suggests the possibility of solving eq.\rf{a.24} by an
iteration
\bel{a.25}
r^{(n+1)} = \delta^{-1} R +\delta^{-1}
\left[ \nabla r^{(n)} + \tfrac{i}{\hbar} \left(r^{(n)}\right)^2\right]
\ee
for $n=0,1,2,\cdots$ with the initial condition $r^{(0)}=0$.
It can be proven \cite{fed} that, for $n\to \infty$,
$r^{(n)}$ converges indeed towards the solution of \rf{a.24}.
The first few iterates are
\bel{a.26}
\bay
r^{(0)} & =  & 0\\
r^{(1)} & =  & \delta^{-1}R\\
r^{(2)} & =  & \delta^{-1}R +\delta^{-1}\nabla(\delta^{-1}R) +O(5)\\
r^{(3)} & =  & r^{(2)} + O(5)
\eay
\ee
Here we observe a property of this iteration which holds true
in general: in order to obtain all terms up to some fixed degree,
only a finite number of iteration steps is needed.
If we are interested only in the terms of degree 4 or less, 
two iterations are sufficient, i.e., 
$r=r^{(2)}+O(5)$. Therefore, with (note that $\nabla_a y^b=0$)
\bel{a.27}
\bay
R &=& \dis\tfrac{1}{4}\, R_{abcd} \,y^c y^d \,d\phi^a d\phi^b \\
\delta^{-1} R &=& \dis-\tfrac{1}{8}\, R_{abcd}\, y^by^c y^d \,d\phi^a\\
\nabla(\delta^{-1}R) &=& -\dis \tfrac{1}{8}\,
\left(\nabla_e R_{abcd}\right)\, y^b y^c y^d \,d\phi^e d\phi^a \\
\delta^{-1}\nabla(\delta^{-1}R)
&=&
\dis\tfrac{1}{40}\, (\nabla_b R_{aecd})\, y^a y^b y^c y^d \,d\phi^e
\eay
\ee
one arrives at eq.\rf{4.30} given in the main text.

A similar method can be used in order to determine the operators
$\co_f(\phi)$ from
\bel{a.28}
\bay
0=D\co_f(\phi) & \equiv & d\co_f(\phi) +\tfrac{i}{\hbar}
                         \left[ \tilde\Gamma, \co_f(\phi)\right]\\[.3cm]
&=& (\nabla -\delta)\,\co_f(\phi) +\tfrac{i}{\hbar}
                         \left[ r, \co_f(\phi)\right]
\eay
\ee
The notation $\co_f(\phi)$ is used also for the symbols
and the correspondence 
$\hat\varphi^a\leftrightarrow y^a$ is understood.
The square brackets in \rf{a.28} denote the commutator
with respect to $\circ$--multiplication. We used that
\bel{a.29}
\tilde\Gamma_a=\Gamma_a +\omega_{ab} \,\hat\varphi^b+r_a
\ee
which implies $D=\nabla-\delta+\tfrac{i}{\hbar}[r, \cdot\,]$.
In section 4
we explained that $\co_f(\phi)$ is a zero--form operator
whose term proportional to the unit operator is given by
the classical observable $f(\phi)$. Hence the projections
\rf{a.7} and \rf{a.8} yield
\bel{a.30}
P_{00}\, \co_f(\phi) =P_0\,\co_f(\phi) =f(\phi)
\ee
Since $\co_f$ is a zero--form, $\delta^{-1} \co_f$ vanishes
identically, and the Hodge decomposition of $\co_f$ reads
therefore $\co_f =f +\delta^{-1}\delta\co_f$.
Thus, every solution of $D\co_f=0$ satisfies
\bel{a.31}
\co_f =f +\delta^{-1}(D+\delta)\co_f
\ee
This equation can be solved by the iteration \cite{fed}
\bel{a.32}
\co_f^{(n+1)} =f +\delta^{-1}(D+\delta)\,\co_f^{(n)}
\quad , \quad n=0,1,2,\cdots
\ee
with $\co_f^{(0)}=f$. Again, the operator $\delta^{-1}(D+\delta)$
increases the degree because $r$ contains only terms of degree 3
and higher. In order to calculate all terms of a fixed
degree, only a finite number of iterations is needed. 
Given $f(\phi)$, the solution of \rf{a.28} is unique;
no subsidiary condition must be imposed. For practical calculations
the component form of \rf{a.32} is more useful:
\bel{a.33}
\co_f^{(n+1)} = f +\cn_y^{-1} y^a\,\nabla_a\,\co_f^{(n)}
+\tfrac{i}{\hbar}\, \cn^{-1}_y \,y^a
\left[r_a ,\co_f^{(n)}\right]
\ee
After three iterations, all terms of degree 3 and less
are stable already: $\co_f=\co_f^{(3)}+O(4)$.
Using the explicit form of $r$, eq.\rf{4.30}, one easily
arrives at eq.\rf{4.31}.

For a flat phase--space with $\Gamma^c_{ab}=0$ and $r=0$
the recurrence relation \rf{a.33} can be solved exactly:
\bel{a.34}
\co_f^{(n)} (\phi) =
\sum_{m=0}^n \frac{1}{m!}\,
\left(y^a\partial_a\right)^m f(\phi)
\ee
Thus, for $n\to\infty$,
\bel{a.35}
\co_f(\phi) =f(\phi+y)
\ee
This important special case is investigated further in section 5.
\end{appendix}

\end{document}